\documentclass[reqno,12pt]{amsart}

\usepackage{rotating} 
\usepackage{amsmath}
\usepackage{amssymb}
\usepackage{amsfonts}
\usepackage{bbm}
\usepackage{graphicx}
\usepackage{epsfig}
\usepackage{epic}
\usepackage[T1]{fontenc}
\newcommand{\nin}{\not\in}
\newcommand{\tmop}[1]{\ensuremath{\operatorname{#1}}}
\newtheorem{theorem}{Theorem}
\newtheorem{definition}{Definition}
\newtheorem{lemma}{Lemma}
\newtheorem{proposition}{Proposition}
\newtheorem{varremark}{Remark}

\numberwithin{equation}{section}
\begin{document}

\title{On the Statistical Mechanics and Surface  Tensions of Binary Mixtures}
\author[  \textsc{J. De Coninck, S. Miracle--Sol\'{e}, and J. Ruiz}
]{
\textsc{J. De Coninck,$^1$ S. Miracle--Sol\'{e},$^2$ and J. Ruiz$^3$}
}

\maketitle

  \parskip 5pt
\begin{quote}

\textsc{\footnotesize Abstract:}
{\footnotesize
Within a lattice model describing a binary mixture with fixed concentrations
of the species we discuss the relation-ship between the surface tension
  of the mixture and the concentrations.}
\\[2mm]
\textsc{\footnotesize Key words:} 
{\footnotesize Surface tensions, binary mixtures, interfaces.}

\end{quote}

\renewcommand{\thefootnote}{}
\footnote{Preprint CPT--2004/P.015, revised, to appear in Journal of Statistical Physics.}
\renewcommand{\thefootnote}{\arabic{footnote}}

\setcounter{footnote}{0} 
\footnotetext[1]{
Centre de Recherche en Mod\'{e}lisation Mol\'{e}culaire, 
Universit\'{e} de Mons--Hainaut, 20 place du Parc, B-7000 Mons, Belgium. 
\hfill\break 
E-mail address: \texttt{Joel.De.Coninck@galileo.umh.ac.be}
} 
\footnotetext[2]{Centre de Physique Th\'{e}orique, CNRS, Luminy case 907, 
F-13288 Marseille Cedex 9, France. 
\hfill\break 
E-mail address: \texttt{Salvador.Miracle-Sole@cpt.univ-mrs.fr}
} 
\footnotetext[3]{Centre de Physique Th\'{e}orique, CNRS, Luminy case 907, 
F-13288 Marseille Cedex 9, France. 
\hfill\break E-mail address: \texttt{Jean.Ruiz@cpt.univ-mrs.fr}
}

\thispagestyle{empty}

\newpage

\section{Introduction}

The notion of surface tension, or interfacial free energy
per unit area, plays a key role in many studies concerning the
surface phenomena and the phase coexistence.

When we consider a solid or a fluid, which is a mixture of two
chemical species 1 and 2, in equilibrium with its vapor,
one of the problems, experimentally as well as theoretically,
is to determine how the corresponding surface tension depends
on the composition of the mixture.

Some relationship is expected which would give this surface tension,
here denoted $\tau_{(1,2)|0}$,
as an interpolation between the two surface tensions,
$\tau_{1|0}$ and $\tau_{2|0}$,
of each of the  species when they are chemically pure.

Using thermodynamical
considerations several equations have been derived in the literature,
according to different assumptions.

Thus, for ideal or nearly ideal solutions,
a fairly simple treatment, due to Guggenheim \cite{Gu}, 
leads to the following equation
\begin{equation}
\label{eq:gug}
   e^{- \beta a^2 \tau_{( 1, 2 ) |0}} = c_1
     e^{- \beta a^2 \tau_{1|0}} + c_2 e^{- \beta a \tau_{2|0}}
\end{equation}
where $c_1$ is
the fixed molar fraction of species $1$ in the $( 1, 2 )$ mixture,
$c_2 = 1 - c_1$, the fixed molar fraction of species $2$,
$a^2$ is the mean surface area per molecule, and $\beta=1/kT$
is the inverse temperature.

A very simple relationship for the so called regular solutions comes from
Prigogine and Defay \cite{DP},
who proposed the equation
\begin{equation}
\label{depr}
\tau_{( 1, 2 ) |0} = c_1 \tau_{1|0} +
     c_2 \tau_{2|0} - K c_1 c_2
\end{equation}
with $K$ a semiempirical constant.

A simple treatment due to Eberhart \cite{Eb}
assumes that the
surface tension of a binary solution is linear in the surface composition,
that is
\begin{equation}
\label{eq:eb}
     \tau_{( 1, 2 ) |0} = c^s_1 \tau_{1|0} + c^s_2 \tau_{2|0}
\end{equation}
where the
$c^s_i$, $i=1,2$, denote the mole fraction near the surface of phase
separation, and that the ratio
$c^s_1 / c_1 $
is proportional to the ratio
$ c_2^s / c_2$.

Finally, when the surface tensions $\tau_{1|0}$ and $\tau_{2|0}$
differ appreciably, a semiempirical equation attributed to Szyszkowsky
(\cite{Sz}, \cite{Sz2})
gives:
\begin{equation}
\label{eq:sz}
\frac{\tau_{( 1, 2 ) |0}}{\tau_{1|0}} = 1 - B
     \ln \left( 1 + \frac{c_2}{A} \right)
\end{equation}
where two characteristic constants $A$ and $B$ of
the compounds have been used,
and $c_2$ is the concentration of the species with the smaller surface
tension.

We refer the reader to Adamson's book \cite{Ad}
(Chapter III, Section 4), and references
therein, for a detailed discussion of the above equations.
On the other hand,
an extensive development for various types of non ideal solutions
that has been made by Defay, Prigogine and co--workers,
can be found in the monography \cite{DPBE}.

More recently, an interface model with a two-valued random interaction was
introduced by two of the present authors in ref.~\cite{DR}
to describe the phase boundary from a microscopic point of view.
The surface tension for that model could be computed
according to a quenched or annealed disorder and one obtains
\begin{eqnarray*}
    \tau^{{\rm quenched}}_{( 1, 2 ), 0} & = & c_1 \tau_{1, 0} + c_2
    \tau_{2, 0}\\
    e^{- \beta \tau^{{\rm annealed}}_{( 1, 2 ) |0}} & = & c_1 e^{- \beta
    \tau_{1, 0}} + c_2 e^{- \beta \tau_{2, 0}}
\end{eqnarray*}
in agreement with the above equations (\ref{eq:gug}) or (\ref{eq:eb}).

The aim of the present paper is two discuss the problem within a
lattice bulk statistical mechanical model describing the binary mixture
in equilibrium with its vapor.
Previous studies of various models of binary lattice gases can be found
in Refs.\ \cite{WW,LG}.

Here,
we consider a lattice gas system with two kinds of particles,
where each lattice site can be in one of the three states, $0,1,2$,
interpreted, respectively, as an empty site,
a site occupied by a particle of the first kind of the model,
and a site occupied by a particle of the second kind.
Whenever the particles $2$ are not allowed the system reduces
to the usual Ising model, in its lattice gas version,
with coupling constant $J_1/2$.
We consider the system in the phase coexistence region and
denote by
$\tau_{1|0}$ the corresponding surface tension
between the dense and the dilute phases.
Analogously, when particles $1$ are not allowed, it reduces to
the Ising model with coupling constant $J_2/2$
and we let $\tau_{2|0}$ be the corresponding surface tension.

We can also study our three state model in the
phase coexistence region (with the help of Pirogov Sinai theory)
and then interpret the dense phase as the binary mixture,
the dilute phase as the corresponding vapor, and
$ \tau_{( 1, 2 ) |0}$ as the surface tension between these two phases.
On the other hand the concentration of particles 1 and 2 in the dense
phase can be fixed to take any given values.

As a main result of this paper we prove that, at low temperatures,
the following equation holds, for the surface tension of our model,
\begin{equation}
e^{- \beta (
     \tau_{( 1, 2 ) |0} - \mathcal{F} )} = c^{\ast}_1 e^{- \beta ( \tau_{1|0} -
     {\mathcal F}_1 )} + c^{\ast}_2 e^{- \beta ( \tau_{2|0} - {\mathcal
F}_2 )_{}}
\end{equation}
   Here
   ${\mathcal{F}}_i$, ($i=1,2$) is
the specific free energy of the  gas of ``jumps'' describing the
Gallavotti's line of phase separation for the Ising model
in two dimensions \cite{G}, and that of the gas of the
``walls'' describing the Dobrushin's microscopic interface \cite{D1}
in three dimensions.
This means that
$\tau_{1|0} - {\mathcal F}_1=J_1$  and $ \tau_{2|0} - {\mathcal F}_2=J_2$
are the respective energy costs
per unit length or unit area of the $1|0$ and the $2|0$ interfaces.
The quantity $\mathcal{F}$ is
the specific free energy
(which can be expressed as a convergent series at low temperatures)
of a gas of some geometrical objects  called aggregates.
In dimension $d=2$, those aggregates are the natural generalizations
to our model of the jumps of Gallavotti's line and the leading term
of the series giving this free energy $\mathcal{F}$ is
\[
- \frac{2}{\beta} \frac{c^*_1 e^{- 2 \beta J_1} + c^*_2 e^{- 2 \beta
     J_2}}{c^*_1 e^{- \beta J_1} + c^*_2 e^{- \beta J_2}}
\]
In dimension $d=3$, they are the natural generalizations of the walls of
the Dobrushin's interface
and then the leading term of the series
is
\[
-\frac{1}{\beta}\frac{  c^*_1 e^{- 5 \beta J_1} + c^*_2 e^{- 5 \beta J_2}
}{
       c^*_1 e^{- \beta J_1} + c^*_2 e^{- \beta J_2}
}
-
\frac{1}{\beta}
\frac{ (c^*_1 e^{- 2 \beta J_1} + c^*_2 e^{- 2 \beta J_2} )^4
}{
(   c^*_1 e^{- \beta J_1} + c^*_2 e^{- \beta J_2}  )^4
}
\]
The coefficients
   $ c^{\ast}_1 $ and $c^{\ast}_2$
 are related to the  concentrations $c_1$ and $c_2$    of the particles $1$ and  the particles $2$
through equation (\ref{eq:ci}) below.
This equation gives at low temperatures:
\begin{eqnarray}
 c^*_i
&=&
c_i
\bigg[
1
-
\big(c_1 e^{-\beta J_1}+c_2 e^{-\beta J_2}\big)^{2d}
-2d c_i e^{-\beta J_i}\big(c_1 e^{-\beta J_1}+c_2 e^{-\beta J_2}\big)^{2d-1} 
\nonumber
\\
&&\hphantom{xxxx} -2(d+1) c_i \big(c_1 e^{-\beta J_1}+c_2 e^{-\beta J_2}\big)^{2d}
+O\big( e^{-(2d+1)\beta \min\{J_1,J_2\}} \big)
\bigg]
\end{eqnarray}
for $i=1,2$.

The paper is organized as follows. The model is defined in Section 2 which also
provides the analysis of the ground states of the system.
Section 3 is devoted to the study of
the Gibbs states of the system at low temperatures and  of
the coexistence between the mixture and the vapor.
Section 4 contains the definitions of the surface tensions and an expansion
of the surface tension between the mixture and the vapor in terms of
interfaces.
Section~5 contains the presentation of the relationship between
surface tensions.
The proofs are given in the two remaining sections.

\section{The model}

We consider a cubic lattice $\mathbb{Z}^d$, of dimension $d=2,3$,
and to each site $\mathbb{Z}^d$ we associate a variable $s_x$
which taking values in the set $\Omega=\{0,1,2\}$ specifies one
of the possible three states of the system at each lattice site.
We say that the site $x$ is empty if $s_x=0$ and that is occupied by
a particle of kind $1$ or of kind $2$ if $s_x=1$ or $2$.
The following Hamiltonian
\begin{equation}
\label{hamiltonien}
H=-\sum_{\langle x,y \rangle}\sum_{\alpha=0}^2\sum_{\beta=0}^2
E_{\alpha\beta}\delta(s_x,\alpha)\delta(s_y,\beta)
\end{equation}
where $\langle x,y \rangle$ denote nearest neighbor pairs, $\delta$ is the usual
Kronecker symbol, $\delta(s,s')=1$ if $s=s'$ and $\delta(s,s')=0$ otherwise,
and $E_{\alpha\beta}=E_{\beta\alpha}$ are the coupling constants,
is of the form of the Blume-Emery-Griffiths model \cite{BEG}
which describes a general three state lattice system for the case of nearest
neighbour interactions.
We shall assume here that
\begin{equation}
\label{condition}
2 E_{12}= E_{11}+E_{22}
\end{equation}
in order to ensure that particles of kinds $1$ and $2$ could be mixed
arbitrarily without any cost of energy.
When these identities are not satisfied two new thermodynamic phases,
either rich in particles of kind 1 or in particles of kind 2,
may appear as equilibrium states of the system.

With the assumption of hypothesis (\ref{condition}) the general Hamiltonian (\ref{hamiltonien})
can be reduced to the form
\begin{eqnarray}
H  =
\sum_{\langle x,y \rangle}
\bigg[
&&
J_1 \big(\delta(s_x,1)    \delta(s_y,0) + \delta(s_x,0)  \delta(s_y,1 ) \big)
\nonumber\\
&&
+
J_2 \big(   \delta(s_x,2)   \delta(s_y,0)+\delta(s_x,0)\delta(s_y,2) \big)
\bigg]
\label{eq:hamilt}
\end{eqnarray}
that is, the case in which the coupling constants satisfy
$E_{10}=E_{01}=-J_1$, $E_{20}=E_{02}=-J_2$ and $E_{\alpha\beta}=0$ otherwise.
Furthermore, we assume that $J_1$ and $J_2$ are positive constants.

In order to see this fact we consider, as it is often convenient,
the reformulation of the three state lattice system in the language of a
magnetic system of spin one.
To do so, we define the spin variable $\sigma_x$ at the $x$ site via
\begin{eqnarray}
\delta(s_x,0) &=&1-\sigma_x^2
\nonumber
\\
\delta(s_x,1) &=& \sigma_x       (\sigma_x + 1 )  / 2
\\
\delta(s_x,2) &=& \sigma_x  ( \sigma_x - 1 ) / 2
\nonumber
\end{eqnarray}
so that $\sigma_x=0,1,-1$ corresponds to the presence at site $x$ of the
state $0,1$ or $2$, respectively.
In terms of the spins, the general Hamiltonian (\ref{hamiltonien}) takes the form
\begin{eqnarray}
H
&=&
\sum_{\langle x, y \rangle}
\mathcal{J}  (\sigma_x-\sigma_y)^2
- \mathcal{K}  \sigma^2_x\sigma^2_y
- \mathcal{C}  (\sigma_x\sigma^2_y+\sigma^2_x\sigma_y)
\nonumber\\
&&
- \sum_{x}(  \mathcal{A}  \sigma_x
+  \mathcal{B }\sigma^2_x)
\label{ghamilt}
\end{eqnarray}
The pair interactions of the system are given
by
\begin{eqnarray}
E_{11}+E_{22}-2E_{12} &=& 8  \mathcal{J}
\nonumber
\\
E_{11}+E_{00}-2E_{01} &=& 2 \mathcal{J} + \mathcal{K} + 2 \mathcal{C}
\label{pairint}
\\
E_{22}+E_{00}-2E_{02} &=& 2  \mathcal{J}  +  \mathcal{K} - 2  \mathcal{C}
\nonumber
\end{eqnarray}
The last two terms can be treated as chemical potentials, with
$$
\mathcal{A} = d(E_{01}-E_{02}) \hbox{ and } \mathcal{B} = d(E_{01}+E_{02}-2E_{00}-2\mathcal{J}) .
$$

We see that our hypothesis (\ref{condition}) implies
\begin{equation}
\label{condj}
\mathcal{J}=0
\end{equation}
On the other hand, taking into account that
\begin{eqnarray*}
2\delta(\sigma_x,1)\delta(\sigma_y,0)
&=&
\sigma_x(\sigma_x+1)(1-\sigma_y^2)
=
\sigma_x^2+\sigma_x-\sigma_x\sigma_y^2-\sigma_x^2\sigma_y^2
\\
2\delta(\sigma_x,-1)\delta(\sigma_y,0)
&=&\sigma_x(\sigma_x-1)(1-\sigma_y^2)
=
\sigma_x^2-\sigma_x+\sigma_x\sigma_y^2-\sigma_x^2\sigma_y^2
\end{eqnarray*}
we obtain Hamiltonian (\ref{eq:hamilt}),  plus chemical potential terms, with
\begin{equation}
2J_1=  K+2C ,
\quad
2J_2=K-2C
\end{equation}

 We notice that condition (\ref{condj}) excludes the models discussed
in the above mentioned refs. \cite{WW} \cite{LG}.
In ref.\ \cite{LG}, Lebowitz and Gallavotti have considered the cases
$2 \mathcal{J}=- \mathcal{K}>0$, $\mathcal{C}=0$, also studied by Wheeler and Widom \cite{WW},
and $\mathcal{J}>0$, $\mathcal{K}=\mathcal{C}=0$, usually known as the Blume-Capel model.
They prove then the appearance at low temperatures of two phases,
respectively reach in particles of kind 1 or of kind 2.
When the two phases coexist one has, as a consequence of the condition
$\mathcal{C}=0$, the equality between the two surface tensions $\tau_{1|0}$ and
$\tau_{2|0}$.
These models, therefore, would not be appropriate for the present study.
The case $\mathcal{C}\ne0$ (and $\mathcal{J}>0$) will be briefly commented in Section 5.

Let us now return to the discussion of Hamiltonian (\ref{eq:hamilt}).
Fixed densities of the three species are introduced through the canonical Gibbs
ensemble of configurations $\mathbf{s}_{\Lambda} =
\left\{ s_x \right\}_{x \in \Lambda}$ in a finite box $\Lambda \subset
\mathbb{Z}^d$, such that
\begin{equation}
  \sum_{x \in \Lambda} \delta ( s_x, 0 ) = N_0, \quad \sum_{x \in \Lambda}
  \delta ( s_x, 1 ) = N_1 \quad \text{and} \quad \sum_{x \in \Lambda} \delta (
  s_x, 2 ) = N_2
\end{equation}
Here $N_0$, $N_1$ and $N_2$ are nonnegative integers satisfying $N_0 + N_1 +
N_2 = | \Lambda |$ where $| \Lambda |$ denotes the number of sites of
$\Lambda$. The associated partition functions with boundary condition bc are
given by
\begin{equation}
  Z_{bc} ( \Lambda ; N_1, N_2 ) = \sum_{\mathbf{s}_{\Lambda} \in
  \Omega^{\Lambda}} e^{- \beta H_{\Lambda} ( \mathbf{s}_{\Lambda} )} \delta
  \left( \sum_{x \in \Lambda} \delta ( s_x, 1), N_1 \right) \delta \left(
  \sum_{x \in \Lambda} \delta ( s_x, 2 ), N_2 \right) \chi^{bc} (
  \mathbf{s}_{\Lambda} )
\end{equation}
where $H_{\Lambda} ( \mathbf{s}_{\Lambda} )$ is the Hamiltonian (\ref{eq:hamilt}) with the sum  over
nearest neighbours pair ¤$\langle x, y \rangle \subset \Lambda$
and
$\chi^{bc} (
\mathbf{s}_{\Lambda} )$ is a characteristic function standing for the boundary
condition bc. We shall be interested in particular to the following boundary
conditions:
\begin{itemize}
  \item the empty boundary condition: $\chi^{\text{emp}} (
  \mathbf{s}_{\Lambda} ) = \prod_{x \in \partial \Lambda} \delta ( s_x, 0 )$
  
  \item the mixture boundary condition: $\chi^{\text{mixt}} (
  \mathbf{s}_{\Lambda} ) = \prod_{x \in \partial \Lambda} ( 1 - \delta ( s_x,
  0 ) )$
  
  \item the free boundary condition: $\chi^{\text{fr}} ( \mathbf{s}_{\Lambda} ) = 1$
\end{itemize}
Hereafter, the boundary $\partial \Lambda$ of the box $\Lambda$ is the set of
sites of $\Lambda$ that have a nearest neighbour in $\Lambda^c = \mathbb{Z}^d
\setminus \Lambda$.

We define the free energy per site corresponding to the above ensemble as a
function of the densities $\rho_1$ and $\rho_2$ of the particles $1$ and $2$:
\begin{equation}
  \label{freeenergy} 
f ( \rho_1, \rho_2 ) = 
\lim_{\Lambda \uparrow \mathbbm{Z}^d} 
-
  \frac{1}{\beta | \Lambda |} 
\ln Z_{bc} ( \Lambda ; [ \rho_1 | \Lambda | ], [\rho_2 | \Lambda | ] )
\end{equation}
where $[\, \cdot \,]$ denotes the integer part and the thermodynamic limit
$\Lambda \uparrow \mathbb{Z}^d$ is taken in the van Hoove 
sense {\cite{R1}}.

We introduce also a grand canonical Gibbs ensemble, which is conjugate to the previous
ensemble, and whose partition function, in the box $\Lambda$ is given by
\begin{equation}
  \label{gcpf} \Xi_{bc} ( \Lambda ; \mu_1, \mu_2 ) =
  \sum_{\mathbf{s}_{\Lambda} \in \Omega^{\Lambda}} e^{- \beta H_{\Lambda} (
  \mathbf{s}_{\Lambda} ) + \mu_1 \sum_{x \in \Lambda} \delta ( s_x, 1 ) +
  \mu_2 \sum_{x \in \Lambda} \delta ( s_x, 2 )}
\end{equation}
where the real numbers $\mu_1$ and $\mu_2$ replace  as 
thermodynamic parameters the densities $\rho_1$ and $\rho_2$. We define the
corresponding specific free energy, the pressure, as the limit
\begin{equation}
  \label{pressure} p ( \mu_1, \mu_2 ) = \lim_{\Lambda \uparrow \mathbbm{Z}^d}
  \frac{1}{| \Lambda |} \ln \Xi_{bc} ( \Lambda ; \mu_1, \mu_2 )
\end{equation}
The equivalence of the two above ensembles is expressed in the following

\begin{theorem}
  Limits (\ref{freeenergy}) and (\ref{pressure}), which define the above free
  energies, exist. They are convex functions of their parameters and are
  related by the Legendre transformations
  \begin{eqnarray}
    p ( \mu_1, \mu_2 ) & = & \sup_{\rho_1, \rho_2} [ \mu_1 \rho_1 + \mu_2 \rho_2 - \beta f
    ( \rho_1, \rho_2 ) ] \\
    \beta f ( \rho_1, \rho_2 ) & = & \sup_{\mu_1, \mu_2} [ \mu_1 \rho_1 + \mu_2 \rho_2 - p
    ( \mu_1, \mu_2 ) ] 
  \end{eqnarray}
  
\end{theorem}

\begin{proof}
  We consider two parallelepipedic boxes $\Lambda'$ and $\Lambda''$ of both
  the same size and paste them to form a parallelepipedic box $\Lambda =
  \Lambda' \cup \Lambda''$ in such a way that $\Lambda' \cap \Lambda'' =
  \emptyset$ and each site of some side of $\Lambda'$ is a nearest neighbour
  of a site of a side of $\Lambda''$. It is easy to see that the following
  sub-additivity property holds:
  \[ Z_{\text{emp}} ( \Lambda ; N'_1 + N''_1, N'_2 + N''_2 ) \geq
     Z_{\text{emp}} ( \Lambda' ; N'_1, N'_2 ) Z_{\text{emp}} ( \Lambda'' ;
     N''_1, N''_2 ) \]
  The same property is shared by the partition function with $\text{mixt}$
  boundary conditions. Then the statements of the theorem follow from standard
  arguments in the theory of the thermodynamic limit {\cite{R1}}.
\end{proof}

We next introduce the finite volume Gibbs measures (a specification)
associated with the second ensemble:
\begin{equation}
\label{eq:gm}
  \mathbb{P}^{bc}_{\Lambda} ( \mathbf{s}_{\Lambda} ) = \frac{e^{- \beta
  \tilde{H}_{\Lambda} ( \mathbf{s}_{\Lambda} )} \chi^{bc} (
  \mathbf{s}_{\Lambda} )}{
\Xi_{bc} ( \Lambda ; \mu_1, \mu_2 )}
\end{equation}
where
\begin{equation}
  \tilde{H}_{\Lambda} ( \mathbf{s}_{\Lambda} ) = H_{\Lambda} (
  \mathbf{s}_{\Lambda} ) - \frac{\mu_1}{\beta} \sum_{x \in \Lambda} \delta (
  s_x, 1 ) - \frac{\mu_2}{\beta} \sum_{x \in \Lambda} \delta ( s_x, 2 )
\end{equation}
They determine by the Dobrushin--Landford--Ruelle equations the set of Gibbs states
$\mathcal{G_{\beta} ( \tilde{H} )}$ on $\mathbb{Z}^d$ corresponding to the
Hamiltonian $\tilde{H}$ at inverse temperature $\beta$ (see e.g. {\cite{R2}}). 
If a Gibbs state $\mathbb{P}
\in \mathcal{G_{\beta} ( \tilde{H} )}$ happens to equal the limit
$\lim_{\Lambda \uparrow \mathbb{Z}^d} \mathbb{P}^{bc}_{\Lambda} ( \cdot )$, we
shall call it the Gibbs state with boundary condition $bc$.

In the zero temperature limit the Gibbs state with empty boundary
condition is concentrated on the configuration with empty sites:
\begin{equation}
  \lim_{\beta \rightarrow \infty} \mathbb{P}^{bc}_{\Lambda} (
  \text{emp}_{\Lambda} ) = 1
\end{equation}
where $\text{emp}_{\Lambda}$ is the configuration where all the sites of
$\Lambda$ are empty, and this limit vanishes for any other configuration.
Gibbs states at $\beta = \infty$ will be called ground
states.

Let
\begin{equation}
  \label{mixtre} R^{_{\text{mixt}}}_{\Lambda} = \left\{ s \in \Omega^{\Lambda}
  : \forall x \in \Lambda, s_x \not= 0 \right\}
\end{equation}
be the \textit{restricted ensemble} of configurations in $\Lambda$ with non
empty sites, and $R^{_{\text{mixt}}}_{\Lambda} ( c )$, $0 \leq c \leq 1$ the
subset of configurations of $R^{_{\text{mixt}}}_{\Lambda}$ with exactly $[ c|
\Lambda | ] = N$ sites occupied by a particle of the specie $1$ (and $|
\Lambda | - [ c| \Lambda | ]$ sites occupied by a particle of the
specie $2$). 
The number of configurations $R^{_{\text{mixt}}}_{\Lambda} ( c )$ equals the
binomial coefficient $\binom{| \Lambda |}{N}$.

With the mixture boundary conditions one has
\begin{equation}
  \label{eq:gsmixt1} 
\lim_{\beta \rightarrow \infty}
  \mathbb{P}^{\text{mixt}}_{\Lambda} ( \mathbf{s}_{\Lambda} ) 
= \frac{e^{\mu_1 [ c| \Lambda | ]}e^{\mu_2 [(1-c)| \Lambda | ]}
}{(  e^{\mu_1}+ e^{\mu_2} )^{| \Lambda |}} \quad
  \text{for{\hspace{0.25em}}each} \quad \mathbf{s}_{\Lambda} \in
  R^{\text{mixt}}_{\Lambda} ( c )
\end{equation}
and
\begin{equation}
  \label{eq:gsmixt2} \lim_{\beta \rightarrow \infty}
  \mathbb{P}^{\text{mixt}}_{\Lambda} ( R^{_{\text{mixt}}}_{\Lambda} ( c ) ) =
  \frac{\binom{| \Lambda |}{N} 
e^{\mu_1 N} 
e^{\mu_2(| \Lambda |- N)}  }{( e^{\mu_1} + e^{\mu_2} )^{| \Lambda |}}
\end{equation}
while this limit vanishes for those 
$\mathbf{s}_{\Lambda} \nin
R^{_{\text{mixt}}}_{\Lambda}$ (the denominator in (\ref{eq:gsmixt1}) and
(\ref{eq:gsmixt2}) is the sum $\sum_{N = 0}^{| \Lambda |}$ of the numerator of
the R.H.S. of (\ref{eq:gsmixt2})). 
Notice that all configurations
$\mathbf{s}_{\Lambda} \in R^{\text{mixt}}_{\Lambda} ( c )$ have the same
probability. 
Moreover, by Stirling's approximation one has for large 
$|\Lambda |$ that 
$\binom{| \Lambda |}{[ c| \Lambda | ]} \approx \left[ \left(
\frac{1}{c} \right)^c \left( \frac{1}{1 - c} \right)^{1 - c} \right]^{|
\Lambda |}$, and the maximum of $e^{\mu_1 c} / (  e^{\mu_1}+ e^{\mu_2} )$ 
is reached
for
\begin{equation}
c = \frac{e^{\mu_1}}{   e^{\mu_1} +  e^{\mu_2}}
\end{equation}
The principle of maximal term gives that, for such values, (\ref{eq:gsmixt2})
tends to $1$ in the thermodynamic limit. This means that the ground state with
$\text{mixt}$ boundary conditions is concentrated on the restricted ensemble
$R^{_{\text{mixt}}} ( c )$ of configurations of non empty sites with
concentration $c$ of particles $1$ and concentration $1 - c$ of particles $2$.

With free boundary conditions, one has
\begin{eqnarray}
  \lim_{\beta \rightarrow \infty} \mathbb{P}^{fr}_{\Lambda} (
  \text{emp}_{\Lambda} ) & = & 
\frac{1}{1 + ( e^{\mu_1} + e^{\mu_2})^{| \Lambda |}} \\
  \lim_{\beta \rightarrow \infty} \mathbb{P}^{fr}_{\Lambda} (
  R^{_{\text{mixt}}}_{\Lambda} ( c ) ) & = & 
\frac{\binom{| \Lambda |}{[ c|
  \Lambda | ]} e^{\mu_1 | \Lambda |}
e^{\mu_2 [(1-c)| \Lambda | ]}}{1 
+ ( e^{\mu_1}+ e^{\mu_2}
  )^{| \Lambda |}} 
\end{eqnarray}
Thus, with the above considerations, we get that for 
$e^{\mu_1}+ e^{\mu_2} =  1$, the configuration with empty 
sites coexists with the restricted
ensemble $R^{_{\text{mixt}}} ( c )$. The diagram of ground states is shown in
Figure 1.

\begin{center}


\setlength{\unitlength}{0.15pt}

\ifx\plotpoint\undefined\newsavebox{\plotpoint}\fi
\sbox{\plotpoint}{\rule[-0.200pt]{0.400pt}{0.400pt}}%


\begin{picture}(1500,900)(0,800)
\font\gnuplot=cmr10 at 10pt
\gnuplot
\sbox{\plotpoint}{\rule[-0.200pt]{0.400pt}{0.400pt}}%

        \drawline(150,859)(1400,859)
        \put(1370,780){$\mu_{1}$}
        \put(1400,842){$\triangleright$}

                \put(950,1092){$R^{\text{mixt}}(c)$}

                \put(350,492){empty}
                \put(250,392){configuration}

        \drawline(800,150)(800,1400)
        \put(680,1400){$\mu_{2}$}
        \put(770,1400){$\triangle$}

\multiput(160.00,848.17)(6.500,-1.000){2}{\rule{1.566pt}{0.400pt}}
\put(173,846.67){\rule{3.132pt}{0.400pt}}
\multiput(173.00,847.17)(6.500,-1.000){2}{\rule{1.566pt}{0.400pt}}
\put(199,845.67){\rule{3.132pt}{0.400pt}}
\multiput(199.00,846.17)(6.500,-1.000){2}{\rule{1.566pt}{0.400pt}}
\put(212,844.67){\rule{3.132pt}{0.400pt}}
\multiput(212.00,845.17)(6.500,-1.000){2}{\rule{1.566pt}{0.400pt}}
\put(225,843.67){\rule{3.132pt}{0.400pt}}
\multiput(225.00,844.17)(6.500,-1.000){2}{\rule{1.566pt}{0.400pt}}
\put(238,842.67){\rule{3.132pt}{0.400pt}}
\multiput(238.00,843.17)(6.500,-1.000){2}{\rule{1.566pt}{0.400pt}}
\put(251,841.67){\rule{2.891pt}{0.400pt}}
\multiput(251.00,842.17)(6.000,-1.000){2}{\rule{1.445pt}{0.400pt}}
\put(263,840.67){\rule{3.132pt}{0.400pt}}
\multiput(263.00,841.17)(6.500,-1.000){2}{\rule{1.566pt}{0.400pt}}
\put(276,839.67){\rule{3.132pt}{0.400pt}}
\multiput(276.00,840.17)(6.500,-1.000){2}{\rule{1.566pt}{0.400pt}}
\put(289,838.67){\rule{3.132pt}{0.400pt}}
\multiput(289.00,839.17)(6.500,-1.000){2}{\rule{1.566pt}{0.400pt}}
\put(302,837.17){\rule{2.700pt}{0.400pt}}
\multiput(302.00,838.17)(7.396,-2.000){2}{\rule{1.350pt}{0.400pt}}
\put(315,835.67){\rule{3.132pt}{0.400pt}}
\multiput(315.00,836.17)(6.500,-1.000){2}{\rule{1.566pt}{0.400pt}}
\put(328,834.17){\rule{2.700pt}{0.400pt}}
\multiput(328.00,835.17)(7.396,-2.000){2}{\rule{1.350pt}{0.400pt}}
\put(341,832.67){\rule{3.132pt}{0.400pt}}
\multiput(341.00,833.17)(6.500,-1.000){2}{\rule{1.566pt}{0.400pt}}
\put(354,831.17){\rule{2.700pt}{0.400pt}}
\multiput(354.00,832.17)(7.396,-2.000){2}{\rule{1.350pt}{0.400pt}}
\put(367,829.17){\rule{2.700pt}{0.400pt}}
\multiput(367.00,830.17)(7.396,-2.000){2}{\rule{1.350pt}{0.400pt}}
\put(380,827.17){\rule{2.700pt}{0.400pt}}
\multiput(380.00,828.17)(7.396,-2.000){2}{\rule{1.350pt}{0.400pt}}
\put(393,825.17){\rule{2.700pt}{0.400pt}}
\multiput(393.00,826.17)(7.396,-2.000){2}{\rule{1.350pt}{0.400pt}}
\multiput(406.00,823.95)(2.695,-0.447){3}{\rule{1.833pt}{0.108pt}}
\multiput(406.00,824.17)(9.195,-3.000){2}{\rule{0.917pt}{0.400pt}}
\put(419,820.17){\rule{2.700pt}{0.400pt}}
\multiput(419.00,821.17)(7.396,-2.000){2}{\rule{1.350pt}{0.400pt}}
\multiput(432.00,818.95)(2.472,-0.447){3}{\rule{1.700pt}{0.108pt}}
\multiput(432.00,819.17)(8.472,-3.000){2}{\rule{0.850pt}{0.400pt}}
\multiput(444.00,815.95)(2.695,-0.447){3}{\rule{1.833pt}{0.108pt}}
\multiput(444.00,816.17)(9.195,-3.000){2}{\rule{0.917pt}{0.400pt}}
\multiput(457.00,812.95)(2.695,-0.447){3}{\rule{1.833pt}{0.108pt}}
\multiput(457.00,813.17)(9.195,-3.000){2}{\rule{0.917pt}{0.400pt}}
\multiput(470.00,809.95)(2.695,-0.447){3}{\rule{1.833pt}{0.108pt}}
\multiput(470.00,810.17)(9.195,-3.000){2}{\rule{0.917pt}{0.400pt}}
\multiput(483.00,806.94)(1.797,-0.468){5}{\rule{1.400pt}{0.113pt}}
\multiput(483.00,807.17)(10.094,-4.000){2}{\rule{0.700pt}{0.400pt}}
\multiput(496.00,802.94)(1.797,-0.468){5}{\rule{1.400pt}{0.113pt}}
\multiput(496.00,803.17)(10.094,-4.000){2}{\rule{0.700pt}{0.400pt}}
\multiput(509.00,798.94)(1.797,-0.468){5}{\rule{1.400pt}{0.113pt}}
\multiput(509.00,799.17)(10.094,-4.000){2}{\rule{0.700pt}{0.400pt}}
\multiput(522.00,794.93)(1.378,-0.477){7}{\rule{1.140pt}{0.115pt}}
\multiput(522.00,795.17)(10.634,-5.000){2}{\rule{0.570pt}{0.400pt}}
\multiput(535.00,789.93)(1.378,-0.477){7}{\rule{1.140pt}{0.115pt}}
\multiput(535.00,790.17)(10.634,-5.000){2}{\rule{0.570pt}{0.400pt}}
\multiput(548.00,784.93)(1.123,-0.482){9}{\rule{0.967pt}{0.116pt}}
\multiput(548.00,785.17)(10.994,-6.000){2}{\rule{0.483pt}{0.400pt}}
\multiput(561.00,778.93)(1.123,-0.482){9}{\rule{0.967pt}{0.116pt}}
\multiput(561.00,779.17)(10.994,-6.000){2}{\rule{0.483pt}{0.400pt}}
\multiput(574.00,772.93)(0.950,-0.485){11}{\rule{0.843pt}{0.117pt}}
\multiput(574.00,773.17)(11.251,-7.000){2}{\rule{0.421pt}{0.400pt}}
\multiput(587.00,765.93)(0.950,-0.485){11}{\rule{0.843pt}{0.117pt}}
\multiput(587.00,766.17)(11.251,-7.000){2}{\rule{0.421pt}{0.400pt}}
\multiput(600.00,758.93)(0.824,-0.488){13}{\rule{0.750pt}{0.117pt}}
\multiput(600.00,759.17)(11.443,-8.000){2}{\rule{0.375pt}{0.400pt}}
\multiput(613.00,750.93)(0.669,-0.489){15}{\rule{0.633pt}{0.118pt}}
\multiput(613.00,751.17)(10.685,-9.000){2}{\rule{0.317pt}{0.400pt}}
\multiput(625.00,741.92)(0.652,-0.491){17}{\rule{0.620pt}{0.118pt}}
\multiput(625.00,742.17)(11.713,-10.000){2}{\rule{0.310pt}{0.400pt}}
\multiput(638.00,731.92)(0.539,-0.492){21}{\rule{0.533pt}{0.119pt}}
\multiput(638.00,732.17)(11.893,-12.000){2}{\rule{0.267pt}{0.400pt}}
\multiput(651.00,719.92)(0.539,-0.492){21}{\rule{0.533pt}{0.119pt}}
\multiput(651.00,720.17)(11.893,-12.000){2}{\rule{0.267pt}{0.400pt}}
\multiput(664.58,706.67)(0.493,-0.576){23}{\rule{0.119pt}{0.562pt}}
\multiput(663.17,707.83)(13.000,-13.834){2}{\rule{0.400pt}{0.281pt}}
\multiput(677.58,691.41)(0.493,-0.655){23}{\rule{0.119pt}{0.623pt}}
\multiput(676.17,692.71)(13.000,-15.707){2}{\rule{0.400pt}{0.312pt}}
\multiput(690.58,674.16)(0.493,-0.734){23}{\rule{0.119pt}{0.685pt}}
\multiput(689.17,675.58)(13.000,-17.579){2}{\rule{0.400pt}{0.342pt}}
\multiput(703.58,654.65)(0.493,-0.893){23}{\rule{0.119pt}{0.808pt}}
\multiput(702.17,656.32)(13.000,-21.324){2}{\rule{0.400pt}{0.404pt}}
\multiput(716.58,631.01)(0.493,-1.091){23}{\rule{0.119pt}{0.962pt}}
\multiput(715.17,633.00)(13.000,-26.004){2}{\rule{0.400pt}{0.481pt}}
\multiput(729.58,602.24)(0.493,-1.329){23}{\rule{0.119pt}{1.146pt}}
\multiput(728.17,604.62)(13.000,-31.621){2}{\rule{0.400pt}{0.573pt}}
\multiput(742.58,566.96)(0.493,-1.726){23}{\rule{0.119pt}{1.454pt}}
\multiput(741.17,569.98)(13.000,-40.982){2}{\rule{0.400pt}{0.727pt}}
\multiput(755.58,520.79)(0.493,-2.400){23}{\rule{0.119pt}{1.977pt}}
\multiput(754.17,524.90)(13.000,-56.897){2}{\rule{0.400pt}{0.988pt}}
\multiput(768.58,455.32)(0.493,-3.787){23}{\rule{0.119pt}{3.054pt}}
\multiput(767.17,461.66)(13.000,-89.662){2}{\rule{0.400pt}{1.527pt}}
\multiput(781.58,344.25)(0.493,-8.466){23}{\rule{0.119pt}{6.685pt}}
\multiput(780.17,358.13)(13.000,-200.126){2}{\rule{0.400pt}{3.342pt}}
\put(186.0,847.0){\rule[-0.200pt]{3.132pt}{0.400pt}}
\end{picture}



\end{center}

\vspace{3.5truecm}
\begin{center}
  \footnotesize{Figure 1: The diagram of ground states.}
\end{center}


In the next section, we extend this analysis to low temperatures.

\section{Coexistence between the mixture and the vapor}

To extend the  analysis of the previous section to the Gibbs states at low
temperatures, we will express the partition functions (\ref{gcpf}) with
empty and $\text{mixt}$ boundary condition in term of contour models.

Let us first introduce the notions of contours by the following definitions.

Consider a configuration $\mathbf{s}_{\Lambda}$ with empty sites on the
boundary $\partial \Lambda$ ($s_x = 0$ for all $x \in \partial \Lambda$). We
define the boundary $B ( \mathbf{s}_{\Lambda} )$ as the set of pairs $\left\{
s_x, s_y \right\}$ such that $s_x \not= 0$ and $s_y = 0$. To a nearest
neighbour pair $\langle x, y \rangle$ let us associate
\begin{description}
  \item[a)] in dimension 2, the unit bond (dual bond) $b_{xy}$ that intersects
  the bond $xy$ in its middle and orthogonal to $xy$.
  \item[b)] in dimension 3, the unit square (dual plaquette) $p_{xy}$ that
  intersects the bond $xy$ in its middle and orthogonal to $xy$.
\end{description}
Two pairs $\left\{ s_x, s_y \right\}$ and $\left\{ s_z, s_t \right\}$ of $B (
\mathbf{s}_{\Lambda} )$ are said adjacent if one of the two conditions is
fulfilled
\begin{description}
  \item[i)] the dual bonds $b_{xy}$ and $b_{zt}$ (respectively the plaquettes
  $p_{xy}$ and $p_{zt}$) are connected.
  \item[ii)] $x = z$, $s_x=s_z\not= 0$, $s_y =s_t =0$, and the bond 
$xy$ with endpoints $x$ and $y$ is parallel
  to the bond with endpoints $z$ and $t$.
\end{description}
A subset $B$ of $B ( \mathbf{s}_{\Lambda} )$ is called connected if the graph
that joins all adjacent pairs of $B$ is connected. It is called contour of the
configuration $\mathbf{s}_{\Lambda}$ if it is a maximal connected  component of 
$B (
\mathbf{s}_{\Lambda} )$.


The boundary and the contours of a configuration $\mathbf{s}_{\Lambda}$ with
occupied sites on the boundary of $\Lambda$ are defined in the same way.

A set $\Gamma$ of pairs $\left\{ s_x, s_y \right\}$ such that $s_x \not= 0$
and $s_y = 0$ is called contour if there exists a configuration
$\mathbf{s}_{\Lambda}$ such that $\Gamma$ is a contour of
$\mathbf{s}_{\Lambda}$. We use $S_{\alpha} ( \Gamma )$ to denote the set of
sites for which $s_x = \alpha$. The set $\text{supp} \hspace{0.25em} \Gamma =
S_0 ( \Gamma ) \cup S_1 ( \Gamma ) \cup S_2 ( \Gamma )$ is called \textit{support} of
the contour $\Gamma$. We will also use $S ( \Gamma ) = S_1 ( \Gamma ) \cup S_2
( \Gamma )$ to denote the set of occupied sites of the contour, $L_1 ( \Gamma
)$ (respectively $L_2 ( \Gamma )$) to denote the number of nearest neighbour
pairs $\langle x, y \rangle$ such that such that $s_x = 1$ and $s_y = 0$
(respectively $s_x = 2$ and $s_y = 0$) and $L ( \Gamma ) = L_1 ( \Gamma ) +
L_2 ( \Gamma )$.

\vspace{-2cm}
\begin{center}

\setlength{\unitlength}{20pt}
\ifx\plotpoint\undefined\newsavebox{\plotpoint}\fi
\begin{picture}(15,15)(0,0)
\font\gnuplot=cmr10 at 10pt
\gnuplot

\put(0,0){$0$}
\put(1,0){$0$}
\put(2,0){$0$}
\put(3,0){$0$}
\put(4,0){$0$}
\put(5,0){$0$}
\put(6,0){$0$}
\put(7,0){$0$}
\put(8,0){$0$}
\put(9,0){$0$}
\put(10,0){$0$}
\put(11,0){$0$}
\put(12,0){$0$}
\put(13,0){$0$}
\put(14,0){$0$}

\put(0,1){$0$}
\put(1,1){$0$}\vspace{5truecm}
\put(2,1){$\mathbf{0}$}
\put(3,1){$\mathbf{0}$}
\put(4,1){$\mathbf{0}$}
\put(5,1){$\mathbf{0}$}
\put(6,1){$\mathbf{0}$}
\put(7,1){$\mathbf{0}$}
\put(8,1){$0$}
\put(9,1){$\mathbf{0}$}
\put(10,1){$\mathbf{0}$}
\put(11,1){$\mathbf{0}$}
\put(12,1){$\mathbf{0}$}
\put(13,1){$0$}
\put(14,1){$0$}

\put(0,2){$0$}
\put(1,2){$\mathbf{0}$}
\put(2,2){$\mathbf{1}$}
\put(3,2){$\mathbf{1}$}
\put(4,2){$\mathbf{2}$}
\put(5,2){$\mathbf{1}$}
\put(6,2){$\mathbf{2}$}
\put(7,2){$\mathbf{1}$}
\put(8,2){$\mathbf{0}$}
\put(9,2){$\mathbf{2}$}
\put(10,2){$\mathbf{1}$}
\put(11,2){$\mathbf{2}$}
\put(12,2){$\mathbf{1}$}
\put(13,2){$\mathbf{0}$}
\put(14,2){$0$}

\put(0,3){$0$}
\put(1,3){$\mathbf{0}$}
\put(2,3){$\mathbf{2}$}
\put(3,3){$1$}
\put(4,3){$2$}
\put(5,3){$1$}
\put(6,3){$2$}
\put(7,3){$\mathbf{1}$}
\put(8,3){$\mathbf{0}$}
\put(9,3){$\mathbf{1}$}
\put(10,3){$2$}
\put(11,3){$2$}
\put(12,3){$\mathbf{2}$}
\put(13,3){$\mathbf{0}$}
\put(14,3){$0$}

\put(0,4){$0$}
\put(1,4){$\mathbf{0}$}
\put(2,4){$\mathbf{1}$}
\put(3,4){$2$}
\put(4,4){$1$}
\put(5,4){$2$}
\put(6,4){$1$}
\put(7,4){$\mathbf{2}$}
\put(8,4){$\mathbf{0}$}
\put(9,4){$\mathbf{2}$}
\put(10,4){$1$}
\put(11,4){$2$}
\put(12,4){$\mathbf{1}$}
\put(13,4){$\mathbf{0}$}
\put(14,4){$0$}

\put(0,5){$0$}
\put(1,5){$\mathbf{0}$}
\put(2,5){$\mathbf{1}$}
\put(3,5){$\mathbf{2}$}
\put(4,5){$\mathbf{1}$}
\put(5,5){$\mathbf{2}$}
\put(6,5){$\mathbf{1}$}
\put(7,5){$\mathbf{2}$}
\put(8,5){$\mathbf{0}$}
\put(9,5){$\mathbf{1}$}
\put(10,5){$\mathbf{2}$}
\put(11,5){$\mathbf{1}$}
\put(12,5){$\mathbf{2}$}
\put(13,5){$\mathbf{0}$}
\put(14,5){$0$}

\put(0,6){$0$}
\put(1,6){$\mathbf{0}$}
\put(2,6){$\mathbf{2}$}
\put(3,6){$\mathbf{0}$}
\put(4,6){$\mathbf{0}$}
\put(5,6){$\mathbf{0}$}
\put(6,6){$\mathbf{0}$}
\put(7,6){$\mathbf{1}$}
\put(8,6){$\mathbf{0}$}
\put(9,6){$\mathbf{0}$}
\put(10,6){$\mathbf{0}$}
\put(11,6){$\mathbf{0}$}
\put(12,6){$\mathbf{0}$}
\put(13,6){$0$}
\put(14,6){$0$}

\put(0,7){$0$}
\put(1,7){$\mathbf{0}$}
\put(2,7){$\mathbf{1}$}
\put(3,7){$\mathbf{0}$}
\put(4,7){$0$}
\put(5,7){$0$}
\put(6,7){$\mathbf{0}$}
\put(7,7){$\mathbf{2}$}
\put(8,7){$\mathbf{0}$}
\put(9,7){$0$}
\put(10,7){$0$}
\put(11,7){$0$}
\put(12,7){$0$}
\put(13,7){$0$}
\put(14,7){$0$}

\put(0,8){$0$}
\put(1,8){$\mathbf{0}$}
\put(2,8){$\mathbf{1}$}
\put(3,8){$\mathbf{0}$}
\put(4,8){$\mathbf{0}$}
\put(5,8){$\mathbf{0}$}
\put(6,8){$\mathbf{0}$}
\put(7,8){$\mathbf{2}$}
\put(8,8){$\mathbf{0}$}
\put(9,8){$0$}
\put(10,8){$0$}
\put(11,8){$0$}
\put(12,8){$0$}
\put(13,8){$0$}
\put(14,8){$0$}

\put(0,9){$0$}
\put(1,9){$\mathbf{0}$}
\put(2,9){$\mathbf{2}$}
\put(3,9){$\mathbf{1}$}
\put(4,9){$\mathbf{2}$}
\put(5,9){$\mathbf{1}$}
\put(6,9){$\mathbf{2}$}
\put(7,9){$\mathbf{2}$}
\put(8,9){$\mathbf{0}$}
\put(9,9){$0$}
\put(10,9){$0$}
\put(11,9){$0$}
\put(12,9){$0$}
\put(13,9){$0$}
\put(14,9){$0$}

\put(0,10){$0$}
\put(1,10){$0$}
\put(2,10){$\mathbf{0}$}
\put(3,10){$\mathbf{0}$}
\put(4,10){$\mathbf{0}$}
\put(5,10){$\mathbf{0}$}
\put(6,10){$\mathbf{0}$}
\put(7,10){$\mathbf{0}$}
\put(8,10){$0$}
\put(9,10){$0$}
\put(10,10){$0$}
\put(11,10){$0$}
\put(12,10){$0$}
\put(13,10){$0$}
\put(14,10){$0$}

\put(0,11){$0$}
\put(1,11){$0$}
\put(2,11){$0$}
\put(3,11){$0$}
\put(4,11){$0$}
\put(5,11){$0$}
\put(6,11){$0$}
\put(7,11){$0$}
\put(8,11){$0$}
\put(9,11){$0$}
\put(10,11){$0$}
\put(11,11){$0$}
\put(12,11){$0$}
\put(13,11){$0$}
\put(14,11){$0$}

\drawline(1.7,1.7)(7.7,1.7)
\drawline(1.7,1.7)(1.7,9.7)
\drawline(1.7,9.7)(7.7,9.7)
\drawline(7.7,1.7)(7.7,9.7)

\drawline(8.7,1.7)(12.7,1.7)
\drawline(8.7,1.7)(8.7,5.7)
\drawline(8.7,5.7)(12.7,5.7)
\drawline(12.7,1.7)(12.7,5.7)

\drawline(2.7,5.7)(6.7,5.7)
\drawline(2.7,5.7)(2.7,8.7)
\drawline(2.7,8.7)(6.7,8.7)
\drawline(6.7,5.7)(6.7,8.7)

\end{picture}
\end{center}

 \vspace{.5 cm}
\begin{center}
 \footnotesize{Figure 2: A configuration with two contours:
the two rectangles on the left belong
to the same contour due to the condition ii) of adjacency.}
\end{center}

Consider the configuration $\mathbf{s}_{\Lambda}$ having $\Gamma$ as unique
contour. The difference $\Lambda \setminus S ( \Gamma )$ splits in components
(set of sites for which the graph that joins all nearest neighbour pairs is
connected) with either all occupied sites or all empty sites. The component
that contains $\partial \Lambda$, denoted $\text{Ext}_{\Lambda}
\hspace{0.25em} \Gamma$ is called \textit{exterior} of the contour. When
$\mathbf{s}_{\Lambda}$ has empty sites (respectively occupied sites) on its
boundary, $\Gamma$ is called $\text{emp}$--contour (respectively
mixt-contour). The \textit{interior} of the contour is the set $\text{Int}
\hspace{0.25em} \Gamma = \Lambda \setminus ( S ( \Gamma ) \cup
\text{Ext}_{\Lambda} \hspace{0.25em} \Gamma )$. It is the union of the
components of $\text{Int}_{\text{emp}} \hspace{0.25em} \Gamma$ with empty
sites with the components of $\text{Int}_{mixt} \hspace{0.25em} \Gamma$ with
occupied sites. Finally $V ( \Gamma ) = S ( \Gamma ) \cup \text{Int}
\hspace{0.25em} \Gamma$.

Two contours $\Gamma_1$ and $\Gamma_2$ are said compatible their union is not
connected. They are mutually compatible external contours if furthermore $V (
\Gamma_1 ) \subset \text{Ext}_{\Lambda \hspace{0.25em}} \Gamma_2$ and $V (
\Gamma_2 ) \subset \text{Ext}_{\Lambda} \hspace{0.25em} \Gamma_1$.

With these definitions one gets the following expansions of the grand canonical
partition functions
\begin{eqnarray}
  \Xi_{\text{emp}} ( \Lambda ; \mu_1, \mu_2 ) 
& =&  
\sum_{\left\{ \Gamma_1, ..., \Gamma_n \right\}_{\text{ext}}}    
  \prod_{i = 1}^n \omega ( \Gamma_i ) \Xi_{\text{emp}} (
  \text{Int}_{\text{emp}} \hspace{0.25em} \Gamma_i ; \mu_1, \mu_2 )
  \Xi_{\text{mixt}} ( \text{Int}_{\text{mixt}} \hspace{0.25em} \Gamma_i ;
  \mu_1, \mu_2 )
\nonumber
\\
&& \label{eq:expvap}\\
  \Xi_{\text{mixt}} ( \Lambda, \mu_1, \mu_2 ) & = & \sum_{\left\{ \Gamma_1,
  ..., \Gamma_n \right\}_{\text{ext}}} (  e^{\mu_1}+e^{\mu_2} )^{| \Lambda \setminus
  \cup_{i = 1}^n V ( \Gamma_i ) |}   \nonumber\\
  &  & \times \prod_{i = 1}^n \omega ( \Gamma_i ) \Xi_{\text{emp}} (
  \text{Int}_{\text{emp}} \hspace{0.25em} \Gamma_i ; \mu_1, \mu_2 )
  \Xi_{\text{mixt}} ( \text{Int}_{\text{mixt}} \hspace{0.25em} \Gamma_i ;
  \mu_1, \mu_2 )\label{eq:expmix}
\end{eqnarray}
where the first sum is over families of mutually external
$\text{emp}$--contours, the second sum is over families of mutually external
$\text{mixt}$--contours, and
\begin{equation}
  \omega ( \Gamma ) = e^{- \beta J_1 L_1 ( \Gamma ) - \beta J_2 L_2 ( \Gamma )
  + \mu_1 |S_1 ( \Gamma ) |+ \mu_2 |S_2 ( \Gamma ) |}
\end{equation}
We put
\begin{equation}
  g_{\text{emp}} = 0, \quad g_{\text{mixt}} = \ln (  e^{\mu_1}+e^{\mu_2} ), \quad
  g_{\max} = \max \{ g_{\text{emp}}, g_{\text{mixt}} \}
\end{equation}
We divide in (\ref{eq:expvap}) each $\Xi_{\text{mixt}}$ by $\Xi_{\text{emp}}$
and multiply it back again in the form (\ref{eq:expvap}). Continuing this
process, and doing an equivalent procedure with (\ref{eq:expmix}), these
relations lead to the following expansion for the partition functions with
boundary condition $q = \text{emp}$ or $q = \text{mixt}$ .
\begin{equation}
  \label{eq:pfcm} \Xi_q ( \Lambda ; \mu_1, \mu_2 ) = e^{g_q | \Lambda |}
  \sum_{_{\{ \Gamma_1, \ldots, \Gamma_n \}_{_{\text{comp}}}}} \prod_{i = 1}^n
  z_q ( \Gamma_i )
\end{equation}
where the sum is now over families of compatible $q$--contours and the
activities $z_q ( \Gamma )$ of contours are given by
\begin{equation}
  \label{eq:activity} z_q ( \Gamma ) = \omega ( \Gamma ) e^{- g_q |S ( \Gamma
  ) |} \frac{\Xi_m ( \text{Int}_m \hspace{0.25em} \Gamma ; \mu_1, \mu_2
  )}{\Xi_q ( \text{Int}_q \hspace{0.25em} \Gamma ; \mu_1, \mu_2 )}
\end{equation}
where $m \not= q$.

The (generalized) Peierls estimates
\begin{equation}
  \label{eq:peierls1} \omega ( \Gamma ) e^{- g_{\max} |S ( \Gamma ) |} \leq
  e^{- \beta JL ( \Gamma )}
\end{equation}
where $J = \min \{ J_1, J_2 \}$,  allow us to have a good control of the
behavior of our system at low enough temperatures using Pirogov-Sinai theory
{\cite{S}}.
Choosing the Zahradnik's formulation of that theory \cite{Z}, we introduce
to state our results,  the following

\begin{definition}
  For $q =  \text{emp}$ and $q= \text{mixt} $ we define the truncated activity
  \[ z'_q ( \Gamma ) = \left\{ \begin{array}{ccc}
       z_q ( \Gamma ) & \text{if} & z_q ( \Gamma ) \leq e^{- \alpha L ( \Gamma
       )}\\
       e^{- \alpha L ( \Gamma )} & \text{otherwise} &
     \end{array} \right. \]
  where $\alpha$ is some positive parameter to be chosen later (see Theorem
  \ref{unicity})
 \end{definition}

\begin{definition}
  The $q$--contour $\Gamma$ is called stable if
  \begin{equation}
    \label{eq:stable} z_q ( \Gamma ) \leq e^{- \alpha L ( \Gamma )}
  \end{equation}
  i.e. if $z_q ( \Gamma ) = z'_q ( \Gamma )$.
  \end{definition}

We define the truncated partition function $\Xi_q' ( \Lambda )$ as the
partition function obtained from (\ref{eq:pfcm}) by leaving out unstable
contours, namely
\begin{equation}
  \label{eq:tpf} \Xi_q' ( \Lambda ;\mu_1,\mu_2) = e^{g_q | \Lambda |} \sum_{_{\{ \Gamma_1,
  \ldots, \Gamma_n \}_{_{\text{comp}}}}} \prod_{i = 1}^n z'_q ( \Gamma_i )
\end{equation}
Here the sum goes over compatible families of \textit{stable $q$--contours}.
Let
\begin{equation}
  \label{eq:thlm} p_q(\mu_1,\mu_2) =  \lim_{\Lambda \uparrow \mathbbm{Z}^d} \frac{1}{|
  \Lambda |} \ln \Xi_q' ( \Lambda;\mu_1,\mu_2 )
\end{equation}
be the \textit{meta-stable pressure} associated with the truncated partition function
\newline
$\Xi_q' ( \Lambda;\mu_1,\mu_2 )$.

For $\alpha$ large enough
the thermodynamic limit
(\ref{eq:thlm}) can be controlled by a convergent cluster expansion.

Namely, to exponentiate the truncated partition function, we
introduce multi-indexes $X$  as functions from the set of contours into the set of non
negative integers (see {\cite{GMM}},{\cite{M}}).
We let $\text{supp} \hspace{0.25em} X = \cup_{\Gamma :
X ( \Gamma ) \geq 1} \text{supp} \hspace{0.25em} \Gamma$ and  define the
truncated functional
\begin{equation}
  \Phi_q ( X ) = \frac{a ( X )}{\prod_{\Gamma} X ( \Gamma ) !} \prod_{\Gamma}
  z_q ( \Gamma )^{X ( \Gamma )}
\end{equation}
where the factor $a ( X )$ is a combinatoric factor defined in terms of the
connectivity properties of the graph $G ( X )$ with vertices corresponding to
$\Gamma$ with $X ( \Gamma ) \geq 1$ (there are $X ( \Gamma )$ vertices for
each such $\Gamma$) that are connected by an edge whenever the corresponding
contours are incompatible). Namely, $a ( X ) = 0$ and hence $\Phi_q ( X ) = 0$
unless $G ( X )$ is a connected graph in which case $X$ is called a
\textit{cluster} and
\begin{equation}
  \label{eq:fcomb} a ( X ) = \sum_{G \subset G ( X )} ( - 1 )^{\left| e ( G )
  \right|}
\end{equation}
Here the sum goes over connected subgraphs $G$ whose vertices coincide with
the vertices of $G ( X )$ and $\left| e ( G ) \right|$ is the number of edges
of the graph $G$. If the cluster $X$ contains only one contour, then 
$a (X ) = 1$.

Note that the number of contours $\Gamma$ with $|S ( \Gamma ) | = s$, $L (
\Gamma ) = n$, whose support contains a given site can be bounded by $2^s
\nu_d^n$ where $\nu_2 = 4$ and $\nu_3 = 14^2$.

As a result of standard cluster expansion we get that for 
$\kappa e^{- \alpha} < 1$ where $\kappa =2 \nu_d \kappa_{\text{cl}}$
and  
$ \kappa_{\text{cl}} \equiv \frac{\sqrt{5} + 3}{2} e^{\frac{2}{\sqrt{5} + 1}}$ 
is the cluster constant \cite{KP}:
\begin{eqnarray}
  \label{eq:B8} \ln
\Xi_q' ( \Lambda;\mu_1,\mu_2 )
& =& 
g_q | \Lambda | + 
\sum_{X : \text{supp} X \subset \Lambda}
\Phi_q ( X )
\\
\label{eq:BB8}
&=&
g_q | \Lambda | + 
| \Lambda |\sum_{X : \text{supp} X \ni x} \frac{\Phi_q ( X )}{| \text{supp}
  \hspace{0.25em} X|} 
+
\sigma (
  \Lambda \mid \Phi_q )
\end{eqnarray}
where
\begin{equation}
  \sigma ( \Lambda \mid \Phi_q ) =  - \sum_{X : \text{supp} X \cap
  \Lambda^c \not= \emptyset} \frac{| \text{supp} X \cap \Lambda |}{|
  \text{supp} \hspace{0.25em} X|} \Phi_q ( X )  \label{eq:bterm}
\end{equation}

In addition (see Section~6):
\begin{eqnarray}
   \sum_{X : \text{supp} X \ni x} \left| \Phi_q ( X
  ) \right|& \leq & \kappa e^{- \alpha} 
\label{eq:BB9}
\\
  \left| \sigma ( \Lambda \mid \Phi_q ) \right| & \leq &
\kappa e^{- \alpha} \left| \partial \Lambda \right|
  \label{eq:B9}
\end{eqnarray}
giving
\begin{equation}
\label{pression}
p_q(\mu_1,\mu_2) = g_q + \sum_{X : \text{supp} \hspace{0.25em} X \ni x} 
 \frac{\Phi_q ( X )}{| \text{supp}
  \hspace{0.25em} X|} 
\end{equation}

The following theorem shows that the low temperature phase diagram of the model is a small
perturbation of the diagram of ground states (see Figure~3).
\begin{theorem}
  \label{unicity}
Assume $\beta$ is large enough so that $e^{- \beta J + 5} =
  e^{- \alpha} < \frac{1}{2 ( d + 1 ) \kappa}$, then there exists a coexistence line
$\ln( e^{\mu_1^*}+e^{\mu_2^*})= O ( e^{- 2d \beta J} )$ on which all mixt--contours and all
$\text{emp}$--contours are stable and such that:
      \begin{equation}
      \label{eq:t1}
 \Xi_q ( \Lambda ;\mu_1^*, \mu_2^*  )= \Xi_{q}' ( \Lambda;\mu_1^*, \mu_2^*   )      
    \end{equation}
for both boundary conditions $q={\rm mixt}$ and $q={\rm emp}$, and
the pressure is given by
\begin{equation}
\label{tt1}
 p(\mu_1^*, \mu_2^* )=p_{\rm mixt}(\mu_1^*, \mu_2^* )=p_{\rm emp}(\mu_1^*, \mu_2^* )
\end{equation}
 For any $t>0$
\begin{equation}
      \label{eq:t2}
 \Xi_{\rm mixt} ( \Lambda ; \mu_1^*+t, \mu_2^* + t )
= \Xi_{\rm mixt}' ( \Lambda;\mu_1^*+t, \mu_2^* + t  ),      
    \end{equation} 
$$p( \mu_1^*+t, \mu_2^* + t)=p_{\rm mixt}( \mu_1^*+t, \mu_2^* + t) >
p_{\rm emp}( \mu_1^*+t, \mu_2^* + t)$$
and
    \begin{equation}
      \label{eq:t3}
 \Xi_{\rm emp}( \Lambda ; \mu_1^*-t, \mu_2^* - t )
= \Xi_{\rm emp}'( \Lambda,  ; \mu_1^*-t, \mu_2^* - t  ),      
    \end{equation}
$$p(  \mu_1^*-t, \mu_2^* - t)=p_{\rm emp}( \mu_1^*-t, \mu_2^* - t)
>p_{\rm mixt}(  \mu_1^*-t, \mu_2^* - t)$$

\end{theorem}

The proof is postponed to Section~6.

\begin{center}

\setlength{\unitlength}{0.15pt}
\ifx\plotpoint\undefined\newsavebox{\plotpoint}\fi
\sbox{\plotpoint}{\rule[-0.200pt]{0.400pt}{0.400pt}}%
\begin{picture}(1500,900)(0,600)
\font\gnuplot=cmr10 at 10pt
\gnuplot
\sbox{\plotpoint}{\rule[-0.200pt]{0.400pt}{0.400pt}}%

        \drawline(150,909)(1400,909)
        \put(1370,830){$\mu_{1}$}
        \put(1400,892){$\triangleright$}

                \put(950,1042){mixture}

                \put(300,492){vapor}

        \drawline(850,150)(850,1400)
        \put(730,1400){$\mu_{2}$}
        \put(820,1400){$\triangle$}

\multiput(160.00,848.17)(6.500,-1.000){2}{\rule{1.566pt}{0.400pt}}
\put(173,846.67){\rule{3.132pt}{0.400pt}}
\multiput(173.00,847.17)(6.500,-1.000){2}{\rule{1.566pt}{0.400pt}}
\put(199,845.67){\rule{3.132pt}{0.400pt}}
\multiput(199.00,846.17)(6.500,-1.000){2}{\rule{1.566pt}{0.400pt}}
\put(212,844.67){\rule{3.132pt}{0.400pt}}
\multiput(212.00,845.17)(6.500,-1.000){2}{\rule{1.566pt}{0.400pt}}
\put(225,843.67){\rule{3.132pt}{0.400pt}}
\multiput(225.00,844.17)(6.500,-1.000){2}{\rule{1.566pt}{0.400pt}}
\put(238,842.67){\rule{3.132pt}{0.400pt}}
\multiput(238.00,843.17)(6.500,-1.000){2}{\rule{1.566pt}{0.400pt}}
\put(251,841.67){\rule{2.891pt}{0.400pt}}
\multiput(251.00,842.17)(6.000,-1.000){2}{\rule{1.445pt}{0.400pt}}
\put(263,840.67){\rule{3.132pt}{0.400pt}}
\multiput(263.00,841.17)(6.500,-1.000){2}{\rule{1.566pt}{0.400pt}}
\put(276,839.67){\rule{3.132pt}{0.400pt}}
\multiput(276.00,840.17)(6.500,-1.000){2}{\rule{1.566pt}{0.400pt}}
\put(289,838.67){\rule{3.132pt}{0.400pt}}
\multiput(289.00,839.17)(6.500,-1.000){2}{\rule{1.566pt}{0.400pt}}
\put(302,837.17){\rule{2.700pt}{0.400pt}}
\multiput(302.00,838.17)(7.396,-2.000){2}{\rule{1.350pt}{0.400pt}}
\put(315,835.67){\rule{3.132pt}{0.400pt}}
\multiput(315.00,836.17)(6.500,-1.000){2}{\rule{1.566pt}{0.400pt}}
\put(328,834.17){\rule{2.700pt}{0.400pt}}
\multiput(328.00,835.17)(7.396,-2.000){2}{\rule{1.350pt}{0.400pt}}
\put(341,832.67){\rule{3.132pt}{0.400pt}}
\multiput(341.00,833.17)(6.500,-1.000){2}{\rule{1.566pt}{0.400pt}}
\put(354,831.17){\rule{2.700pt}{0.400pt}}
\multiput(354.00,832.17)(7.396,-2.000){2}{\rule{1.350pt}{0.400pt}}
\put(367,829.17){\rule{2.700pt}{0.400pt}}
\multiput(367.00,830.17)(7.396,-2.000){2}{\rule{1.350pt}{0.400pt}}
\put(380,827.17){\rule{2.700pt}{0.400pt}}
\multiput(380.00,828.17)(7.396,-2.000){2}{\rule{1.350pt}{0.400pt}}
\put(393,825.17){\rule{2.700pt}{0.400pt}}
\multiput(393.00,826.17)(7.396,-2.000){2}{\rule{1.350pt}{0.400pt}}
\multiput(406.00,823.95)(2.695,-0.447){3}{\rule{1.833pt}{0.108pt}}
\multiput(406.00,824.17)(9.195,-3.000){2}{\rule{0.917pt}{0.400pt}}
\put(419,820.17){\rule{2.700pt}{0.400pt}}
\multiput(419.00,821.17)(7.396,-2.000){2}{\rule{1.350pt}{0.400pt}}
\multiput(432.00,818.95)(2.472,-0.447){3}{\rule{1.700pt}{0.108pt}}
\multiput(432.00,819.17)(8.472,-3.000){2}{\rule{0.850pt}{0.400pt}}
\multiput(444.00,815.95)(2.695,-0.447){3}{\rule{1.833pt}{0.108pt}}
\multiput(444.00,816.17)(9.195,-3.000){2}{\rule{0.917pt}{0.400pt}}
\multiput(457.00,812.95)(2.695,-0.447){3}{\rule{1.833pt}{0.108pt}}
\multiput(457.00,813.17)(9.195,-3.000){2}{\rule{0.917pt}{0.400pt}}
\multiput(470.00,809.95)(2.695,-0.447){3}{\rule{1.833pt}{0.108pt}}
\multiput(470.00,810.17)(9.195,-3.000){2}{\rule{0.917pt}{0.400pt}}
\multiput(483.00,806.94)(1.797,-0.468){5}{\rule{1.400pt}{0.113pt}}
\multiput(483.00,807.17)(10.094,-4.000){2}{\rule{0.700pt}{0.400pt}}
\multiput(496.00,802.94)(1.797,-0.468){5}{\rule{1.400pt}{0.113pt}}
\multiput(496.00,803.17)(10.094,-4.000){2}{\rule{0.700pt}{0.400pt}}
\multiput(509.00,798.94)(1.797,-0.468){5}{\rule{1.400pt}{0.113pt}}
\multiput(509.00,799.17)(10.094,-4.000){2}{\rule{0.700pt}{0.400pt}}
\multiput(522.00,794.93)(1.378,-0.477){7}{\rule{1.140pt}{0.115pt}}
\multiput(522.00,795.17)(10.634,-5.000){2}{\rule{0.570pt}{0.400pt}}
\multiput(535.00,789.93)(1.378,-0.477){7}{\rule{1.140pt}{0.115pt}}
\multiput(535.00,790.17)(10.634,-5.000){2}{\rule{0.570pt}{0.400pt}}
\multiput(548.00,784.93)(1.123,-0.482){9}{\rule{0.967pt}{0.116pt}}
\multiput(548.00,785.17)(10.994,-6.000){2}{\rule{0.483pt}{0.400pt}}
\multiput(561.00,778.93)(1.123,-0.482){9}{\rule{0.967pt}{0.116pt}}
\multiput(561.00,779.17)(10.994,-6.000){2}{\rule{0.483pt}{0.400pt}}
\multiput(574.00,772.93)(0.950,-0.485){11}{\rule{0.843pt}{0.117pt}}
\multiput(574.00,773.17)(11.251,-7.000){2}{\rule{0.421pt}{0.400pt}}
\multiput(587.00,765.93)(0.950,-0.485){11}{\rule{0.843pt}{0.117pt}}
\multiput(587.00,766.17)(11.251,-7.000){2}{\rule{0.421pt}{0.400pt}}
\multiput(600.00,758.93)(0.824,-0.488){13}{\rule{0.750pt}{0.117pt}}
\multiput(600.00,759.17)(11.443,-8.000){2}{\rule{0.375pt}{0.400pt}}
\multiput(613.00,750.93)(0.669,-0.489){15}{\rule{0.633pt}{0.118pt}}
\multiput(613.00,751.17)(10.685,-9.000){2}{\rule{0.317pt}{0.400pt}}
\multiput(625.00,741.92)(0.652,-0.491){17}{\rule{0.620pt}{0.118pt}}
\multiput(625.00,742.17)(11.713,-10.000){2}{\rule{0.310pt}{0.400pt}}
\multiput(638.00,731.92)(0.539,-0.492){21}{\rule{0.533pt}{0.119pt}}
\multiput(638.00,732.17)(11.893,-12.000){2}{\rule{0.267pt}{0.400pt}}
\multiput(651.00,719.92)(0.539,-0.492){21}{\rule{0.533pt}{0.119pt}}
\multiput(651.00,720.17)(11.893,-12.000){2}{\rule{0.267pt}{0.400pt}}
\multiput(664.58,706.67)(0.493,-0.576){23}{\rule{0.119pt}{0.562pt}}
\multiput(663.17,707.83)(13.000,-13.834){2}{\rule{0.400pt}{0.281pt}}
\multiput(677.58,691.41)(0.493,-0.655){23}{\rule{0.119pt}{0.623pt}}
\multiput(676.17,692.71)(13.000,-15.707){2}{\rule{0.400pt}{0.312pt}}
\multiput(690.58,674.16)(0.493,-0.734){23}{\rule{0.119pt}{0.685pt}}
\multiput(689.17,675.58)(13.000,-17.579){2}{\rule{0.400pt}{0.342pt}}
\multiput(703.58,654.65)(0.493,-0.893){23}{\rule{0.119pt}{0.808pt}}
\multiput(702.17,656.32)(13.000,-21.324){2}{\rule{0.400pt}{0.404pt}}
\multiput(716.58,631.01)(0.493,-1.091){23}{\rule{0.119pt}{0.962pt}}
\multiput(715.17,633.00)(13.000,-26.004){2}{\rule{0.400pt}{0.481pt}}
\multiput(729.58,602.24)(0.493,-1.329){23}{\rule{0.119pt}{1.146pt}}
\multiput(728.17,604.62)(13.000,-31.621){2}{\rule{0.400pt}{0.573pt}}
\multiput(742.58,566.96)(0.493,-1.726){23}{\rule{0.119pt}{1.454pt}}
\multiput(741.17,569.98)(13.000,-40.982){2}{\rule{0.400pt}{0.727pt}}
\multiput(755.58,520.79)(0.493,-2.400){23}{\rule{0.119pt}{1.977pt}}
\multiput(754.17,524.90)(13.000,-56.897){2}{\rule{0.400pt}{0.988pt}}
\multiput(768.58,455.32)(0.493,-3.787){23}{\rule{0.119pt}{3.054pt}}
\multiput(767.17,461.66)(13.000,-89.662){2}{\rule{0.400pt}{1.527pt}}
\multiput(781.58,344.25)(0.493,-8.466){23}{\rule{0.119pt}{6.685pt}}
\multiput(780.17,358.13)(13.000,-200.126){2}{\rule{0.400pt}{3.342pt}}
\put(186.0,847.0){\rule[-0.200pt]{3.132pt}{0.400pt}}
\end{picture}
\end{center}

\vspace{3cm}
\begin{center}
 \footnotesize{Figure 3: Phase diagram at low temperature.}
\end{center}

Whenever a cluster $X$ contains only one contour $\Gamma$ ($X (
\Gamma ) = 1$ and $X ( \Gamma' ) = 0$ for $\Gamma' \neq \Gamma$) one has
$\Phi_q ( X ) = z_q ( \Gamma )$. 
From this property we get that the metastable
pressures read:
\begin{eqnarray}
  p_{\text{mixt}} ( \mu_1, \mu_2 )
&=& \ln ( e^{\mu_1} + e^{\mu_2} )
 + \frac{(
  e^{\mu_{1 - \beta J_1}} + e^{\mu_2 - \beta J_2} )^{2d}}{( e^{\mu_1} + e^{\mu_2}
  )^{2d+1}}  + O \big( e^{- (2d+1) \beta J} \big)   
\label{pmix}
\\
  p_{\text{emp}} ( \mu_1, \mu_2 )
&= &e^{\mu_1 - 2d \beta J_1} + e^{\mu_2 - 2d
  \beta J_2} + O \big( e^{- (2d+1) \beta J} \big) 
\label{pemp}
\end{eqnarray}
Equalizing (\ref{pmix}) with (\ref{pemp}) gives the first term
of the equation for the coexistence line stated in Theorem \ref{unicity}.

Let us introduce the infinite volume expectation $\langle\, \cdot \,
\rangle^{\text{bc}} ( \mu_1, \mu_2 )$ associated to the Gibbs measure
(\ref{eq:gm}):
\begin{equation}
  \langle \, \cdot \, \rangle^{\text{bc}} ( \mu_1, \mu_2 ) 
= \lim_{\Lambda \uparrow
  \mathbbm{Z}^d} \sum_{\mathbf{s}_{\Lambda} \in \Omega^{\Lambda}} \cdot\,
  \mathbbm{P}_{\Lambda}^{\text{bc}} (\mathbf{s}_{\Lambda} )
\end{equation}
 As a consequence of the cluster expansion we have
for any $t \geq 0$:
\begin{eqnarray}
  \langle \delta ( s_x, 1 ) + \delta ( s_x, 2 ) \rangle^{\text{mixt}} 
(\mu_1^{*} + t, \mu_2^* + t )
&=& 
\langle 1 - \delta ( s_x, 0 )
  \rangle^{\text{mixt}} ( \mu_1^{*} + t, \mu_2^* + t )
\nonumber
\\ 
&\geq&  1 - O \big( e^{- 2d \beta J} \big)
 \label{c1} \\
  \langle \delta ( s_x, 1 ) + \delta ( s_x, 2 ) \rangle^{\text{emp}} (
  \mu_1^{*} - t, \mu_2^* - t )& =& \langle 1 - \delta ( s_x, 0 )
  \rangle^{\text{emp}} ( \mu_1^{*} - t, \mu_2^* - t ) 
\nonumber
\\
&\leq &   O \big( e^{- 2d \beta J}\big)
  \label{c2} 
\end{eqnarray}
This shows that the model exhibits at low temperature a first
order phase transition at the coexistence line where the pressure is
discontinuous. 

Let, 
$c_i
=\frac{\partial p_{\text{mixt}} ( \mu_1, \mu_2 )}{ \partial \mu_i}\vert_{\mu_1=\mu^*_1,\mu_2=\mu^*_2}$, 
$i=1$, $2$ 
be the density of the particle $i$ in the mixture regime, on the coexistence line,
and let
\begin{eqnarray}
d_i
&=&\frac{\partial }{ \partial \mu_i}
\sum_{X : \text{supp} \hspace{0.25em} X \ni x} 
 \frac{\Phi_{\text{mixt}} ( X )}{| \text{supp}
  \hspace{0.25em} X|} 
\bigg\vert_{\mu_1=\mu^*_1,\mu_2=\mu^*_2}
\nonumber
\\
&=&
2d e^{\mu_i^* - \beta J_i}
\frac{
(
  e^{\mu_1^* - \beta J_1} + e^{\mu_2^* - \beta J_2} )^{2d-1}}{( e^{\mu_1^*} + e^{\mu_2^*}
  )^{2d+1}}
\nonumber
\\& & - 
(2d+1) e^{\mu_i^* }
\frac{
(
  e^{\mu_1^* - \beta J_1} + e^{\mu_2^* - \beta J_2} )^{2d}}{( e^{\mu_1^*} + e^{\mu_2^*}
  )^{2d+2}}
+
 O \big( e^{- (2d+1) \beta J} \big)
\nonumber
\end{eqnarray} 
for $i=1$, $2$. 
One has
\begin{equation}
\label{cmix}
  c_i
=\frac{e^{ \mu_i^*}  }{  e^{\mu_1^*}+e^{ \mu_2^*}  }+d_i
=
\frac{
e^{ \mu_i^*   }    }{    e^{ \mu_1^*}+e^{ \mu_2^*}  }
(1+
(e^{ \mu_1^*}+e^{ \mu_2^*})d_i )
=\frac{
e^{ \mu_i^*   }    }{    e^{ \mu_1^*}+e^{ \mu_2^*}  }
\big(1+ O \big( e^{- 2d \beta J} \big)\big)
\end{equation}

\section{Surface tensions}

To introduce the surface tension between the mixture and the vapor, we
consider the parallelepipedic box:
\[ V = V_{L, M} = \left\{ ( x_1, .., x_d ) \in \mathbb{Z}^d : |x_i | \leq L, i
   = 1, ..., d - 1 ; - M \leq x_d \leq M - 1 \right\} \]
Let $\partial_+ V$ (respectively $\partial_- V$) be the set of sites of
$\partial V$ with $x_d \geq 0$ (respectively $x_d < 0$). We introduce the
boundary condition
\[ \chi^{\text{mixt}, \text{emp}} ( \mathbf{s}_{V} ) = \prod_{x \in
   \partial_- V} ( 1 - \delta ( s_x, 0 ) ) \prod_{x \in \partial_+ V} \delta (
   s_x, 0 ) \]
This boundary condition enforces the existence of an interface (see below for
its precise definition) between the mixture and the vapor. The interfacial
tension between the mixture and the vapor is defined by the limit
\begin{equation}
  \label{eq:surtenmel} \tau_{\text{mixt}, \text{emp}} = \lim_{L \rightarrow
  \infty} \lim_{M \rightarrow \infty} \frac{F ( V )}{( 2 L + 1 )^{d - 1}}
\end{equation}
where
\begin{equation}
  \label{eq:Fen} F ( V ) = - \frac{1}{\beta} \ln \frac{\Xi_{\text{mixt},
  \text{emp}} ( V ; \mu^*_1, \mu^*_2 )}{( \Xi_{\text{mixt}} ( V ;\mu^*_1, \mu^*_2  ) 
\Xi_{\text{emp}} ( V ;\mu^*_1, \mu^*_2 ) )^{1 / 2}}
\end{equation}
This definition is justified by noticing that in this expression the volume
terms proportional to the free energy of the coexisting phases, as well as the
terms corresponding to the boundary effects, cancel and only the term that
takes into account the free energy of the interface is left.

To give a precise description of interfaces, we let 
$\mathbb{L}_+$ denotes the semi infinite lattice with $x_d \geq 0$ and let
$\mathbb{L}_- = \mathbb{Z}^d \setminus \mathbb{L}_+$ denotes its complement.
Consider then a configuration $\mathbf{s} \in \Omega^{\mathbb{Z}^d}$, with empty
sites on $\partial_+ V$ and on $\mathbb{L}_+ \setminus V$ and with occupied sites
on a $\partial_- V$ and on $\mathbb{L}_- \setminus V$. The boundary $B (
\mathbf{s} )$ (set of pairs $\left\{ s_x, s_y \right\}$ such that $s_x \not=
0$ and $s_y = 0$) of such configuration necessarily contains an infinite
component $I_{\infty}$ whose support outside the box $V$ is the set of n.n.\
pairs between $\mathbb{L}_+$ and $\mathbb{L}_-$.

We call interface $I$ the part of $I_{\infty}$ whose support lies inside the
box $V$: $I_{\infty} \setminus I$, is called extension of $I$. As it was done
for contours, we use $S_{\alpha} ( I )$ to denote the set of sites for which
$s_x = \alpha$. The set $\text{supp} \hspace{0.25em} I = S_0 ( I ) \cup S_1 (
I ) \cup S_2 ( I )$ is called support of the interface $I$. We will also use
$S ( I ) = S_1 ( I ) \cup S_2 ( I )$ to denote the set of occupied sites of
the interface, $L_1 ( I )$ (respectively $L_2 ( I )$) to denote the number of
nearest neighbour pairs $\langle x, y \rangle$ such that such that $s_x = 1$
and $s_y = 0$ (respectively $s_x = 2$ and $s_y = 0$) and $L ( I ) = L_1 ( I )
+ L_2 ( I )$.

\vspace{2cm}
\begin{center}
\setlength{\unitlength}{20pt}

\ifx\plotpoint\undefined\newsavebox{\plotpoint}\fi
\begin{picture}(15,9)(0,0)
\font\gnuplot=cmr10 at 10pt
\gnuplot

\put(0,0){$2$}
\put(1,0){$1$}
\put(2,0){$2$}
\put(3,0){$2$}
\put(4,0){$2$}
\put(5,0){$1$}
\put(6,0){$2$}
\put(7,0){$1$}
\put(8,0){$2$}
\put(9,0){$1$}
\put(10,0){$1$}
\put(11,0){$1$}
\put(12,0){$2$}
\put(13,0){$2$}

\put(0,1){$2$}
\put(1,1){$1$}
\put(2,1){$1$}
\put(3,1){$1$}
\put(4,1){$1$}
\put(5,1){$2$}
\put(6,1){$2$}
\put(7,1){$1$}
\put(8,1){$2$}
\put(9,1){$2$}
\put(10,1){$2$}
\put(11,1){$2$}
\put(12,1){$2$}
\put(13,1){$1$}

\put(0,2){$1$}
\put(1,2){$2$}
\put(2,2){$0$}
\put(3,2){$1$}
\put(4,2){$1$}
\put(5,2){$2$}
\put(6,2){$1$}
\put(7,2){$2$}
\put(8,2){$0$}
\put(9,2){$0$}
\put(10,2){$2$}
\put(11,2){$1$}
\put(12,2){$1$}
\put(13,2){$1$}

\put(0,3){$2$}
\put(1,3){$1$}
\put(2,3){$1$}
\put(3,3){$2$}
\put(4,3){$1$}
\put(5,3){$1$}
\put(6,3){$\mathbf{1}$}
\put(7,3){$1$}
\put(8,3){$0$}
\put(9,3){$0$}
\put(10,3){$2$}
\put(11,3){$2$}
\put(12,3){$1$}
\put(13,3){$2$}

\put(0,4){$2$}
\put(1,4){$2$}
\put(2,4){$1$}
\put(3,4){$1$}
\put(4,4){$2$}
\put(5,4){$\mathbf{1}$}
\put(6,4){$\mathbf{0}$}
\put(7,4){$\mathbf{2}$}
\put(8,4){$2$}
\put(9,4){$2$}
\put(10,4){$1$}
\put(11,4){$1$}
\put(12,4){$1$}
\put(13,4){$1$}

\put(0,5){$\mathbf{1}$}
\put(1,5){$\mathbf{2}$}
\put(2,5){$\mathbf{1}$}
\put(3,5){$2$}
\put(4,5){$2$}
\put(5,5){$\mathbf{1}$}
\put(6,5){$\mathbf{0}$}
\put(7,5){$\mathbf{2}$}
\put(8,5){$2$}
\put(9,5){$2$}
\put(10,5){$2$}
\put(11,5){$1$}
\put(12,5){$\mathbf{2}$}
\put(13,5){$\mathbf{1}$}

\put(0,6){$\mathbf{0}$}
\put(1,6){$\mathbf{0}$}
\put(2,6){$\mathbf{0}$}
\put(3,6){$\mathbf{1}$}
\put(4,6){$2$}
\put(5,6){$\mathbf{2}$}
\put(6,6){$\mathbf{0}$}
\put(7,6){$\mathbf{1}$}
\put(8,6){$\mathbf{2}$}
\put(9,6){$\mathbf{1}$}
\put(10,6){$\mathbf{2}$}
\put(11,6){$\mathbf{1}$}
\put(12,6){$\mathbf{0}$}
\put(13,6){$\mathbf{0}$}

\put(0,7){$0$}
\put(1,7){$\mathbf{0}$}
\put(2,7){$\mathbf{0}$}
\put(3,7){$\mathbf{1}$}
\put(4,7){$2$}
\put(5,7){$\mathbf{2}$}
\put(6,7){$\mathbf{0}$}
\put(7,7){$\mathbf{0}$}
\put(8,7){$\mathbf{0}$}
\put(9,7){$\mathbf{0}$}
\put(10,7){$\mathbf{0}$}
\put(11,7){$\mathbf{0}$}
\put(12,7){$0$}
\put(13,7){$0$}

\put(0,8){$\mathbf{0}$}
\put(1,8){$\mathbf{2}$}
\put(2,8){$\mathbf{1}$}
\put(3,8){$\mathbf{1}$}
\put(4,8){$\mathbf{1}$}
\put(5,8){$\mathbf{2}$}
\put(6,8){$\mathbf{0}$}
\put(7,8){$0$}
\put(8,8){$0$}
\put(9,8){$0$}
\put(10,8){$0$}
\put(11,8){$0$}
\put(12,8){$0$}
\put(13,8){$0$}

\put(0,9){$0$}
\put(1,9){$\mathbf{0}$}
\put(2,9){$\mathbf{0}$}
\put(3,9){$\mathbf{0}$}
\put(4,9){$\mathbf{0}$}
\put(5,9){$\mathbf{0}$}
\put(6,9){$0$}
\put(7,9){$0$}
\put(8,9){$1$}
\put(9,9){$2$}
\put(10,9){$1$}
\put(11,9){$0$}
\put(12,9){$0$}
\put(13,9){$0$}

\put(0,10){$0$}
\put(1,10){$0$}
\put(2,10){$0$}
\put(3,10){$0$}
\put(4,10){$0$}
\put(5,10){$0$}
\put(6,10){$0$}
\put(7,10){$0$}
\put(8,10){$1$}
\put(9,10){$2$}
\put(10,10){$0$}
\put(11,10){$0$}
\put(12,10){$0$}
\put(13,10){$0$}

\put(0,11){$0$}
\put(1,11){$0$}
\put(2,11){$0$}
\put(3,11){$0$}
\put(4,11){$0$}
\put(5,11){$0$}
\put(6,11){$0$}
\put(7,11){$0$}
\put(8,11){$0$}
\put(9,11){$0$}
\put(10,11){$0$}
\put(11,11){$0$}
\put(12,11){$0$}
\put(13,11){$0$}

\drawline(-0.3,5.7)(2.7,5.7)
\drawline(2.7,5.7)(2.7,7.7)
\drawline(2.7,7.7)(0.7,7.7)

\drawline(0.7,7.7)(0.7,8.7)
\drawline(0.7,8.7)(5.7,8.7)

\drawline(5.7,8.7)(5.7,3.7)
\drawline(5.7,3.7)(6.7,3.7)

\drawline(6.7,3.7)(6.7,6.7)
\drawline(6.7,6.7)(11.7,6.7)

\drawline(11.7,6.7)(11.7,5.7)
\drawline(11.7,5.7)(13.7,5.7)

\dashline{.1}(7.7,8.7)(10.7,8.7)
\dashline{.1}(10.7,8.7)(10.7,9.7)

\dashline{.1}(10.7,9.7)(9.7,9.7)
\dashline{.1}(9.7,9.7)(9.7,10.7)

\dashline{.1}(9.7,10.7)(7.7,10.7)
\dashline{.1}(7.7,10.7)(7.7,8.7)
\dashline{.1}(7.7,1.7)(9.7,1.7)
\dashline{.1}(9.7,1.7)(9.7,3.7)
\dashline{.1}(9.7,3.7)(7.7,3.7)
\dashline{.1}(7.7,3.7)(7.7,1.7)
\dashline{.1}(1.7,1.7)(2.7,1.7)
\dashline{.1}(2.7,1.7)(2.7,2.7)
\dashline{.1}(2.7,2.7)(1.7,2.7)
\dashline{.1}(1.7,2.7)(1.7,1.7)

\end{picture}
\end{center}
\vspace{.5cm}
\begin{center}
\footnotesize{Figure~4: Interface of a configuration with the $\text{mixt},\text{emp}$
boundary condition.}
\end{center}

The set $V \setminus S ( I )$ splits into a part $D$ below the interface and a
part $U$ above the interface: if one consider the configuration whose boundary
contains only the interface $I$, $D$ is the subset of $V \setminus S ( I )$
with occupied sites and $U$ is the subset of $V \setminus S ( I )$ with empty
sites.

Then, the partition function, with mixed boundary conditions, can be expanded
over interfaces as follows
\
\begin{equation}
   \Xi_{\text{mixt}, \text{emp}} ( V ;\mu^*_1, \mu^*_2 ) = \sum_{I :
   \text{supp} \hspace{0.25em} I \subset V} \omega ( I ) \Xi_{\text{mixt}} ( D
   ;\mu^*_1, \mu^*_2 ) \Xi_{\text{emp}} ( U ;\mu^*_1, \mu^*_2 ) 
\end{equation}
where
\begin{equation}
 \omega ( I ) = e^{- \beta J_1 L_1 ( I ) - \beta J_2 L_2 ( I ) + \mu_1^* |S_1
   ( I ) |+ \mu_2^* |S_2( I ) |} 
\end{equation}
Therefore,
\begin{equation}
  \label{eq:actst} e^{- \beta F ( V )} = \sum_{I : \text{supp} \hspace{0.25em}
  I \subset V} \omega ( I ) \frac{\Xi_{\text{mixt}} ( D ;\mu^*_1, \mu^*_2 )
  \Xi_{\text{emp}} ( U ;\mu^*_1, \mu^*_2 )}{( \Xi_{\text{mixt}} ( V ;\mu^*_1, \mu^*_2 ) 
\Xi_{\text{emp}} ( V ;\mu^*_1, \mu^*_2 ) )^{1 / 2}}
\end{equation}
Since $\Xi_q ( \Lambda ;\mu^*_1, \mu^*_2 ) = \Xi'_q ( \Lambda;\mu^*_1, \mu^*_2 )$ 
for both empty and mixt boundary conditions, we get by
(\ref{eq:BB8} ) and (\ref{pression})
\[ 
\Xi_p ( \Lambda ; \mu^*_1, \mu^*_2 )
= \exp \left\{ p_q | \Lambda |
   + \sigma ( \Lambda \mid \Phi_q ) \right\}
\]
Applying this formula to the various partition functions of (\ref{eq:actst}),
and taking into account equations (\ref{tt1}), we get:
\begin{equation}
  \label{eq:z} e^{- \beta F ( V )} = \sum_{I : \text{supp} \hspace{0.25em} I
  \subset V} \omega ( I ) e^{-p_(\mu_1^*,\mu_2^*) |S ( I ) | + \sigma ( D \mid
  \Phi_{\text{mixt}} ) + \sigma ( U \mid \Phi_{\text{emp}} ) 
- \frac{1}{2} \sigma ( V \mid \Phi_{\text{mixt}} )
- \frac{1}{2} \sigma ( V \mid \Phi_{\text{emp}} ) 
}
\end{equation}
Let $V_+^c = ( \mathbb{Z}^d \setminus V ) \cap \mathbb{L}_+$ and $V_-^c = (
\mathbb{Z}^d \setminus V ) \cap \mathbb{L}_-$. We then apply the formula
(\ref{eq:bterm}) to the four last terms of the exponent of the RHS of
(\ref{eq:z}) and rewrite the different contributions according to appropriated
and natural decompositions of the sets over which the sums take place and
collecting analogous terms. Before applying this formula it is convenient first to
 sum over the clusters with same support. Thus we let
\begin{eqnarray*}
  \tilde{\Phi}_{\text{mixt}} ( C ) & = & \sum_{X : \text{supp} \hspace{0.25em}
  X = C} \Phi_{\text{mixt}} ( X )\\
  \tilde{\Phi}_{\text{emp}} ( C ) & = & \sum_{X : \text{supp} \hspace{0.25em}
  X = C} \Phi_{\text{emp}} ( X )
\end{eqnarray*}
We then get
\begin{equation}
\label{eq:fv}
 e^{- \beta F ( V )} = e^{- K_V} \sum_{I : \text{supp} \hspace{0.25em} I
   \subset V} \omega ( I ) e^{-p(\mu_1^*,\mu_2^*) |S ( I ) |} A ( I ) B_V ( I ) 
\end{equation}
where
\begin{equation}
\label{eq:kk}
K_V = \frac{1}{2} \sum_{C : C \cap V_+^c \not= \emptyset \atop C \cap V_-^c
   \not= \emptyset} \tilde{\Phi}_{\text{mixt}} ( C ) \frac{|C \cap V|}{|C|}
   + \frac{1}{2} \sum_{C : C \cap V_+^c \not= \emptyset \atop C \cap V_-^c
   \not= \emptyset} \tilde{\Phi}_{\text{emp}} ( C ) \frac{|C \cap V|}{|C|}
\end{equation}
\begin{equation}
\label{eq:aa}
 A ( I ) = \prod_{C : C \cap S ( I ) \not= \emptyset} e^{-
   \tilde{\Phi}_{\text{mixt}} ( C ) \frac{|C \cap D|}{|C|}} \prod_{C : C \cap
   S ( I ) \not= \emptyset} e^{- \tilde{\Phi}_{\text{emp}} ( C ) \frac{|C
   \cap U|}{|C|}} 
\end{equation}
and
\begin{eqnarray}
  B_V ( I ) & = & \prod_{C : C \cap S ( I ) \not= \emptyset \atop C \cap V_-^c
  \not= \emptyset} e^{\tilde{\Phi}_{\text{mixt}} ( C ) \frac{|C \cap
  D|}{|C|}} \prod_{C : C \cap V_-^c \not= \emptyset}
  e^{\tilde{\Phi}_{\text{mixt}} ( C ) \frac{|C \cap ( V \setminus D )
  |}{|C|}}
\nonumber
\\
  &  & \prod_{C : C \cap S ( I ) \not= \emptyset \atop C \cap V_+^c \not=
  \emptyset} e^{\tilde{\Phi}_{\text{emp}} ( C ) \frac{|C \cap U|}{|C|}}
  \prod_{C : C \cap V_+^c \not= \emptyset} e^{\tilde{\Phi}_{\text{emp}} ( C
  ) \frac{|C \cap ( V \setminus U) |}{|C|}}
\label{eq:bb}
\end{eqnarray}
In these formula, it is understood that the arguments $C$ of
$\tilde{\Phi}_{\text{mixt}}$ and $\tilde{\Phi}_{\text{emp}}$ are respectively
supports of clusters of $\text{mixt}$--contours and clusters of
$\text{emp}$--contours. These functionals satisfy the bound:
\begin{equation}
  \label{eq:borneclusterm} | \tilde{\Phi} ( C ) | \leq L ( C ) ( \kappa e^{-
  \alpha} )^{L ( C )}
\end{equation}
where $L ( C )$ is the number of n.n.\  pairs of $C$ (see Section~6).

This  property allows us  to prove that the limit of $F(V)$ when
$M \to \infty$ exists, and that, if we denote by
$ \overline{V}$ the infinite cylinder $\lim_{M \to \infty}
V_{L, M}$, then one gets actually 
$\lim_{M\to \infty}F(V)\newline =F( \overline{V})$ with $F( \overline{V})$
defined as in (\ref{eq:fv})--(\ref{eq:bb}).

The surface tension then reads
\begin{equation}
  \label{eq:st}
  \tau_{\text{mixt}, \text{emp}} = \lim_{L \rightarrow
  \infty} \frac{F (  \overline{V} )}{( 2 L + 1 )^{d - 1}}\end{equation}

Clearly the term 
$K_{\overline{V}} /
\beta ( 2 L + 1 )^{d - 1}$ 
tends to $0$ in the limit $L \rightarrow \infty$.
Let us introduce the modified free energy
\begin{equation}
  \label{eq:freeen} e^{- \beta F' (  \overline{V} )} = \sum_{I : \text{supp}
  \hspace{0.25em} I \subset  \overline{V}} \omega ( I ) e^{-p(\mu_1^*,\mu_2^*) |S ( I ) |} A ( I )
\end{equation}
As a consequence of the analysis of Section~7 we shall see that
 the free energy  $F'(\overline{V})$
differs from $F(\overline{V})$ only by a term proportional to $L^{d-2}$, 
thus showing  that the surface tension 
is also  given by (\ref{eq:st}) with  $F(\overline{V})$ replaced by  $F'(\overline{V})$.
To simplify notations, we shall only consider this last free energy.

In the Solid-On-Solid (SOS) approximation, that we will also consider, the
surface tension reads
\begin{equation}
  \label{eq:surtenmelsos} \tau^{\text{SOS}}_{\text{mixt}, \text{emp}} =
  \lim_{L \rightarrow \infty} \frac{F^{\text{SOS}} ( \overline{V} )}{( 2 L + 1
  )^{d - 1}}
\end{equation}
where
\[ 
F^{\text{SOS}} ( \overline{V} ) 
= - \frac{1}{\beta} 
\ln
   \sum_{I^{\text{SOS}} : \text{supp} \hspace{0.25em} I \subset \overline{V}}
   \omega ( I ) (  e^{\mu_1^*}+e^{\mu_2^*} )^{- |S ( I ) |} 
\]
where the SOS interfaces belonging to $I^{\text{SOS}}$ have no overhangs. This
means that the set of dual bonds (in $2$--dimensions) or dual plaquettes 
(in $3$--dimensions) of such  interface corresponds to the graph of a function.
This approximation may be obtained by adding to the Hamiltonian 
(\ref{eq:hamilt})  the  anisotropic interaction
\[
 \sideset{}{^{\text{vert}}} \sum_{\langle x, y \rangle }
 J' \left[ \delta ( s_x, 0 ) ( 1 -
  \delta ( s_y, 0 ) ) + ( 1 - \delta ( s_x, 0 ) ) \delta ( s_y, 0 ) \right]
\]
where the sum $\sideset{}{^{\text{vert}}} \sum$ is over vertical bonds,
and then taking, with an appropriated normalization, the
limit $J' \rightarrow \infty$.
The coexistence line in that approximation coincides with the ground states coexistence line,
$e^{\mu_1^*}+e^{\mu_2^*}=1$, so that:
\begin{equation}
  \label{eq:freeensos} e^{- \beta F^{\text{SOS}} ( \overline{V} ) } = \sum_{I^{\text{SOS}} : \text{supp}
  \hspace{0.25em} I \subset  \overline{V} } \omega ( I ) 
\end{equation}

Let us now define the surface tensions between each species of the mixture and
the vapor. 
As mentioned in the introduction, 
whenever either the particles $1$ 
or the particles $2$ are not allowed the system reduces 
to the usual Ising model in its lattice gas version.  
Thus,  we introduce the configurations 
$n_{V} \in
\left\{ 0, 1 \right\}^{V}$ 
of the lattice gas and the following
partition functions
\begin{eqnarray*}
  Q_{\alpha} ( V ) 
& = & 
\sum_{n_{V} \in \left\{ 0, 1\right\}^{V}} 
\exp 
\left\{ 
\beta J_{\alpha} 
\sum_{\langle x, y \rangle  \subset V} 
[n_x ( 1 - n_y )+ (1-n_x) n_y] 
\right\} \prod_{x \in \partial V}
  n_x
\\
  Q_{\alpha, 0} ( V ) 
& = & 
\sum_{n_{V} \in \left\{ 0, 1 \right\}^{V}} 
\exp 
\left\{ 
\beta J_{\alpha} \sum_{\langle x, y \rangle
  \subset V} 
[n_x ( 1 - n_y )+ (1-n_x) n_y]
\right\} \\
&&
\hphantom{xxxxxxxxxxxxxxxxxxxxxxxxxx}
\times     \prod_{x \in \partial_- V} (
  1 - n_x ) \prod_{x \in \partial_+ V} n_x
\end{eqnarray*}
for $\alpha = 1$ and $\alpha = 2$.

The interfacial tension between the species $\alpha=1,2$, 
and the vapor is the limit (\cite{GMM,BLP2})
\[
\tau_{\alpha, 0} = \lim_{L \rightarrow \infty} \lim_{M
   \rightarrow \infty} \frac{F_{\alpha} ( V )}{( 2 L + 1 )^{d - 1}}
\]
where
\[ F_{\alpha} ( V ) = - \frac{1}{\beta} \ln \frac{Q_{\alpha, 0} ( V
   )}{Q_{\alpha} ( V )}
\]
It is well known that the ratio $Q_{\alpha, 0} ( V ) / Q_{\alpha} ( V )$ can
be expressed as a sum over interfaces which in this case are connected set of
bonds or plaquettes of the dual lattice \cite{G,D1}. Extracting the energy of the flat
interface, the system can be written as a gas of excitations leading to
\[ F_{\alpha} ( V ) = J_{\alpha} ( 2 L + 1 )^{d - 1} + F_{\alpha}^{\text{ex}}
   ( V ) \]
In two dimensions, $F_{\alpha}^{\text{ex}}$ is the free energy  of  the gas of jumps
of the Gallavotti's line \cite{G}.
In three dimensions,  $F_{\alpha}^{\text{ex}}$ is the free energy of the gas  of walls
of the Dobrushin's interface \cite{D1}.
In both cases these free energies can be analyzed by cluster expansion techniques
at low temperatures. Namely,  the specific free energies
$\mathcal{F}_{\alpha} = \lim_{L
\rightarrow \infty} F_{\alpha}^{\text{ext}} ( V ) / ( 2 L + 1 )^{d - 1}$ 
exist and are given by convergent expansions in term of the activities
$e^{-\beta J_{\alpha}}$,  giving
\begin{equation}
\label{eq:tjf}
  \beta \tau_{\alpha, 0} = \beta J_{\alpha} + \beta \mathcal{F}_{\alpha}
\end{equation}
In addition
\begin{eqnarray}
  -\beta \mathcal{F}_{\alpha} & = & 2 e^{- \beta J_{\alpha}} + O ( e^{- 2 \beta
  J_{\alpha}} ) \quad \text{for} \quad d = 2  \label{eq:stim2}\\
  -\beta \mathcal{F}_{\alpha} & = & 2 e^{- 4 \beta J_{\alpha}} + O ( e^{- 6
  \beta J_{\alpha}} ) \quad \text{for} \quad d = 3  \label{eq:stim3}
\end{eqnarray}

We refer the reader also to  \cite{BLP}, \cite{Mi}, \cite{A} and
references therein for the study of these expansions. 
Furthermore,  in two dimensions the surface tension defined above is known to coincide with 
the one computed by Onsager \cite{GMM}. 
We thus have an exact expression for $\tau_{\alpha, 0}$, 
and  for  $\mathcal{F}_{\alpha}$:
\begin{equation}
  \label{eq:ed2} \beta \mathcal{F}_{\alpha} =  \ln \tanh( \beta J_{\alpha}/2)
\end{equation}
for $\beta J_{\alpha}$ larger than the critical value $ \ln (1+\sqrt{2})$.

Similar results hold in the corresponding SOS approximation (see {\cite{P}}
for the three $3$--dimensional case).
We will use 
$\tau^{\text{SOS}}_{\alpha, \text{emp}}$
 and 
$\mathcal{F}^{\text{SOS}}_{\alpha}$ 
the surface
tensions and free energies in this approximation.
In two dimensions 
$\mathcal{F}^{\text{SOS}}_{\alpha} $  
is also
exactly known \cite{Temp}  and also given by (\ref{eq:ed2}), but for all temperatures:
\begin{equation}
  \label{eq:ed2s} \beta \mathcal{F}^{\text{SOS}}_{\alpha} =  \ln \tanh( \beta J_{\alpha}/2)
\end{equation}
for $\beta\geq 0$.

\section{Main Results}

In this section we will give a relationship between the surface tensions
introduced in the previous section.

The leading term of the free energy $F' ( \overline{V} )$
corresponds to flat interfaces without any decoration. 
They are those interfaces, for which the set of
n.n.\
$x, y$ such that $x$ is empty and $y$ is occupied crosses the plane $x_d
= - 1 / 2$, and such that $A(I)=1$. 
Let $\mathcal{I}_{\text{flat}}$ be the set of flat interfaces and
$N = ( 2 L + 1 )^{d - 1}$. We have 
\begin{eqnarray*}
  e^{- \beta F_{\text{flat}} ( \overline{V} )}
& \equiv & 
\sum_{I :  \text{supp} \hspace{0.25em} I \subset \overline{V}
\atop 
I \in  \mathcal{I}_{\text{flat}}   } 
\omega ( I ) 
e^{- p ( \mu_1^{\ast},  \mu_2^{\ast} ) |S ( I ) |}
\\
  & = & 
\sum_{n = 0}^N \binom{N}{n} 
e^{( \mu_1^{\ast} - \beta J_1 ) n}
  e^{( \mu_1^{\ast} - \beta J_2 ) ( N - n )} 
e^{- p ( \mu_1^{\ast}, \mu_2^{\ast} ) N}
\\
  & = & ( c^*_1 e^{-\beta  J_1} + c^*_2 e^{-\beta J_2} )^N
\end{eqnarray*}
where 
\begin{equation}
\label{concent}
c^*_1 = e^{ \mu_1^{\ast} - p (\mu_1^{\ast}, \mu_2^{\ast} )}
, \quad
c^*_2 = e^{\mu_2^{\ast}
- 
p ( \mu_1^{\ast}, \mu_2^{\ast} )}
\end{equation}

We will show in Section~7 that the difference 
$F' ( \overline{V} )- F_{\text{flat}} ( \overline{V} )$ 
can be expressed as a free energy of a gas of excitations called aggregates.
It will then turns out that the limit
\begin{equation}
\label{ffff}
\mathcal{F} = \lim_{L\to\infty} 
\frac{F' ( \overline{V} )- F_{\text{flat}} ( \overline{V} )}{N}
\end{equation}
exists and is given by a convergent expansion at low temperatures,
see (\ref{eq:series2}).   

\begin{theorem}
\label{T:gen}
Assume $\beta$ is large enough so that
$ 8 e(e-1) \kappa^2 e^{- \beta J + 5}<1$, then 
the interfacial tensions
  $\tau_{\rm mixt, emp}$, $\tau_{1,0}$ and $\tau_{2,0}$ 
satisfy the equation:
\begin{equation}
\label{eq:loi} 
e^{- \beta (
     \tau_{\rm mixt, emp} - \mathcal{F} )} = c^*_1 e^{- \beta (
     \tau_{1, 0} - \mathcal{F}_1 )} 
+ c^*_2 e^{- \beta ( \tau_{2,0} - \mathcal{F}_2 )} 
\end{equation}
where 
\begin{equation}
c^*_i 
= e^{ \mu_i^* - p (\mu_1^*, \mu_2^* )}
\end{equation}
$\mathcal{F}$ is the convergent series defined by
  (\ref{eq:series2}):
\begin{equation}\label{eq:free1} 
-\beta \mathcal{F} = \frac{c^*_1 e^{- 5 \beta J_1} + c^*_2 e^{-
     5 \beta J_2}}{c^*_1 e^{- \beta J_1} + c^*_2 e^{- \beta J_2}} + \frac{( c^*_1
     e^{- 2 \beta J_1} + c^*_2 e^{- 2 \beta J_2} )^4}{( c^*_1 e^{- \beta J_1} +
     c^*_2 e^{- \beta J_2} )^4} + O ( e^{- 5 \beta J} ) 
\end{equation}
in dimension $d = 3$
  and
\begin{equation}\label{eq:free2} 
-\beta \mathcal{F} = 2 \frac{c^*_1 e^{- 2 \beta J_1} + c^*_2 e^{- 2 \beta
     J_2}}{c^*_1 e^{- \beta J_1} + c^*_2 e^{- \beta J_2}} + O ( e^{- 2 \beta J} )
  \end{equation}
in dimension $d = 2$, and the convergent series $\mathcal{F}_{\alpha}$
satisfies (\ref{eq:stim2})(\ref{eq:stim3}) and (\ref{eq:ed2}).
\end{theorem}
The proof is postponed to Section~7.

Some remarks and comments are in order.

The  densities $c_i$ are easily related to
the $c^*_i$ through the relation (see (\ref{pmix}), (\ref{cmix})):
\begin{eqnarray}
c_i^*
&=&
c_i
\frac{
(e^{ \mu_1^*} + e^{ \mu_2^*})
e^{  -p_{\text{mixt}}(\mu^*_1, \mu^*_2)   }
}{ 1 + (e^{ \mu_1^*} + e^{  \mu_2^*})   d_i
}
\label{eq:ci}
\\
&=&
c_i
\bigg[
1
-
\big(c_1 e^{-\beta J_1}+c_2 e^{-\beta J_2}\big)^{2d}
-2d c_i e^{-\beta J_i}\big(c_1 e^{-\beta J_1}+c_2 e^{-\beta J_2}\big)^{2d-1}
\nonumber
\\
&&\hphantom{xxxx} -2(d+1) c_i \big(c_1 e^{-\beta J_1}+c_2 e^{-\beta J_2}\big)^{2d}
+O\big( e^{-(2d+1)\beta J   } \big)
\bigg]
\end{eqnarray}

In the SOS approximation, as mentioned in the previous section, 
the coexistence line is 
given by $e^{\mu^*_1}+e^{\mu^*_1}=1$ 
and the densities of each species in the mixture regime on this line are
$c_1=e^{\mu^*_1}$ and $c_2=e^{\mu^*_2}$. We have 
$$
  e^{- \beta F_{\text{flat}}^{\text{SOS}} ( \overline{V} )}
 \equiv  
\sum_{I :  \text{supp} \hspace{0.25em} I \subset \overline{V}
\atop 
I \in  \mathcal{I}_{\text{flat}}   } 
\omega ( I ) 
 = ( c_1 e^{-\beta  J_1} + c_2 e^{-\beta J_2} )^N
$$
and the equation between the surface tensions reads
\begin{equation}
  \label{eq:sos}
  e^{- \beta ( \tau^{\text{SOS}}_{\text{mixt}, \text{emp}} -
     \mathcal{F}^{\text{SOS}} )} 
= c_1 e^{- \beta ( \tau^{\text{SOS}}_{1, 0} -
     \mathcal{F}^{\text{SOS}}_1 )} 
+ c_2 e^{- \beta ( \tau^{\text{SOS}}_{2, 0} -
     \mathcal{F}^{\text{SOS}}_2 )} 
\end{equation}
Here 
$\mathcal{F}^{\text{SOS}}=
\lim_{L\to\infty} 
\frac{F^{\text{SOS}} ( \overline{V} )- F_{\text{flat}}^{\text{SOS}} ( \overline{V} )}{N}
$: it can be expressed as  the series given by 
(\ref{eq:series2}), but with 
truncated functional corresponding to the activities 
$$
z^{\text{SOS}}(W)=e^{-
   \beta J_1 L_1 ( W ) - \beta J_2 L_2 ( W ) + \mu_1^{\ast} |S_1 ( W ) | +
   \mu_2^{\ast} |S_2 ( W ) |}$$
   and satisfy also relations (\ref{eq:free1}) and (\ref{eq:free2}) with $c^*_i$ replaced by $c_i$.



To find an exact solution for $\mathcal{F}^{SOS}$ in two dimensions 
seems to be an interesting problem.
Actually, this would give an exact equation for 
the surface tensions not restricted to low temperatures \cite{Progress}.


The method developed here    can be  extended naturally to finite range interactions.
As an example, we can consider the Hamiltonian (\ref{eq:hamilt})  with the sum over nearest
neighbours and next nearest neighbours. In that case we found that the corresponding surface tensions satisfy, in two dimensions and at low enough temperatures, the equation
\begin{eqnarray*}
e^{- \beta (
     \tau_{\rm mixt, emp} - \mathcal{F} )}
     &=&
      c^*_1 e^{- 3 \beta J_1 }
+ c^*_2 e^{-3 \beta J_2 }
\\
 &=&
      c^*_1 e^{- \beta (
     \tau_{1, 0} - \mathcal{F}_1 )}
+ c^*_2 e^{- \beta ( \tau_{2,0} - \mathcal{F}_2 )}
\end{eqnarray*}
where
\[
-\beta \mathcal{F}_1 = 2 e^{- \beta J_1}  +O(e^{- 2\beta J_1}) ,
\quad
-\beta \mathcal{F}_2 = 2 e^{- \beta J_2} +O(e^{- 2\beta J_2})
\]
and
\begin{eqnarray*}
  - \beta \mathcal{F} & = & \frac{c_1^{\ast 3} e^{- 7 \beta J_1} + c_1^{\ast
  2} c_2^{\ast} ( e^{- 6 \beta J_1 - J_2} + e^{- 5 J_1 - 2 J_2} + e^{- 4 J_1 -
  3 J_2} )}{( c_1^{\ast} e^{- 3 \beta J_1} + c_2^{\ast} e^{- 3 \beta J_1}
  )^2}\\
  &  & + \frac{c_1^{\ast} c_2^{\ast 2} ( e^{- 3 \beta J_1 - 4 J_2} + e^{- 2
  J_1 - 5 J_2} + e^{- J_1 - 7 J_2} ) + c_2^{\ast 3} e^{- 7 \beta J_2}}{(
  c_1^{\ast} e^{- 3 \beta J_1} + c_2^{\ast} e^{- 3 \beta J_1} )^2}  +  O(e^{- 2\beta J})
\end{eqnarray*}
with $c^*_i = c_i \big(1+O(e^{- 8 \beta J})\big)$.

When the hypothesis (\ref{condition}) or (\ref{condj}) on the model defined by Hamiltonian
(\ref{hamiltonien}) or (\ref{ghamilt}) is not satisfied, with
 $\mathcal{J}>0$,   there are, at low temperatures,
 two phases rich in particles of species $1$
 or of species $2$ as been proved in \cite{LG} for $\mathcal{C}=0$.
 This can be also shown for
 $\mathcal{C}\not= 0$ provided that the three quantities in equations (\ref{pairint}) are positive.
 Indeed the Hamiltonian (\ref{hamiltonien}) and  (\ref{ghamilt}) can be reduced, up to chemical potential
 terms, to the form
  \begin{eqnarray}
H  =
\sum_{\langle x,y \rangle}
\bigg[
&&
J_1 \big(\delta(s_x,1)    \delta(s_y,0) + \delta(s_x,0)  \delta(s_y,1 ) \big)
\nonumber\\
&&
+
J_2 \big(   \delta(s_x,2)   \delta(s_y,0)+\delta(s_x,0)\delta(s_y,2) \big)
\bigg]
\nonumber\\
&&
+
 \frac{\mathcal{J}}{4} \big(   \delta(s_x,1)   \delta(s_y,2)+\delta(s_x,2)\delta(s_y,1) \big)
\bigg]
\end{eqnarray}
where $J_1= 2 \mathcal{J} + \mathcal{K} + 2\mathcal{C}$,   $J_1= 2 \mathcal{J} + \mathcal{K} - 2\mathcal{C}$,
and $ \mathcal{J}$ are positive constants.
  In this situation and with fixed concentration of the species either we are
 in one of these two pure phases that coexist with the gaseous phase or in the coexistence of
 the three phases.
 In the later case, there will be segregation between the three phases and
 one will observe only the three kinds of interfaces  between
 these phases.
  One cannot truly speak in this approach about a real surface
 tension of a mixture.

Let us stress that from our equations (\ref{eq:loi}) and (\ref{eq:sos}), we get obviously the
Guggeinheim relation (\ref{eq:gug}) by neglecting the terms exponentially small with
the inverse temperature, i.e. $\mathcal{F},\mathcal{F}_1,$ and
$\mathcal{F}_2$.
This shows that this relation provides a good approximation
at very low temperatures.

To see the quantitative difference between the above equations, let us
consider a concrete example.
Note that when we take only into account the (exponentially) smallest order
corrections  in our equations, they reads (in dimension $3$):
\begin{eqnarray}
  \tau_{\tmop{mixt}, \tmop{emp}}
   & = &
   - \frac{k T}{a^2} 
   \ln 
   \left[ 
   c_1 
   e^{-
  \frac{a^2 \tau_{1, 0}}{k T} 
  - 2 e^{- \frac{4 a^2 \tau_{1, 0}}{k T}}} 
  + 
  ( 1 - c_1 ) e^{- \frac{a^2 \tau_{2, 0}}{k T} 
  - 2 e^{- \frac{4 a^2 \tau_{2, 0}}{k T}}
  } 
  \right]
  \label{tauapprox}
  \\
  & + & 
   k T 
  \frac{c_1 e^{- \frac{5 a^2 \tau_{1, 0}}{k T}} + ( 1 - c_1 ) e^{-
  \frac{5 a^2 \tau_{2, 0}}{k T}}}{c_1 e^{- \frac{a^2 \tau_{1, 0}}{k T}} + ( 1
  - c_1 ) 
  e^{- \frac{a^2 \tau_{2, 0}}{k T}}}
  + k T 
  \left[ 
  \frac{c_1 e^{- \frac{2
  a^2 \tau_{1, 0}}{k T}} 
  + ( 1 - c_1 ) e^{- \frac{2 a^2 \tau_{2, 0}}{k
  T}}}{c_1 e^{- \frac{a^2 \tau_{1, 0}}{k T}} + ( 1 - c_1 ) e^{- \frac{a^2
  \tau_{2, 0}}{k T}}} 
  \right]^4
  \nonumber
\end{eqnarray}
where $a$ is the lattice spacing and $k$ the Boltzman constant, while for the
Guggenheim relation, one has
\begin{equation}
\tau^{\tmop{Gugg}}_{\tmop{mixt}, \tmop{emp}} = - \frac{k T}{a^2}
\ln 
\left[ 
c_1
   e^{- \frac{a^2 \tau_{1, 0}}{k T}} + ( 1 - c_1 )  e^{- \frac{a^2 \tau_{2,
   0}}{k T}} 
   \right]
    \label{taugugg}
   \end{equation}
   Note also that the simplest form of equations  (\ref{depr}) and (\ref{eq:eb}) reads
\begin{equation}
\label{taudfe}
\tau_{( 1, 2 ) |0} = c_1 \tau_{1,0} + (1- c_1) \tau_{2,0}
\end{equation}
Figure~5 shows that there is actually a quantitative difference
between the  equations (\ref{tauapprox}) and  (\ref{taugugg}). We have chosen a mixture of water and hexane, i.e.\
$\tau_{1, 0} = 0.0724 N / m$, $\tau_{2, 0} = 0.01978 N / m$,
$a = 3 \overset{\; \circ}{A}$, the size of a water molecule, and the temperature $T=300 K$.


\begin{rotate}{-90}
\epsfig{file=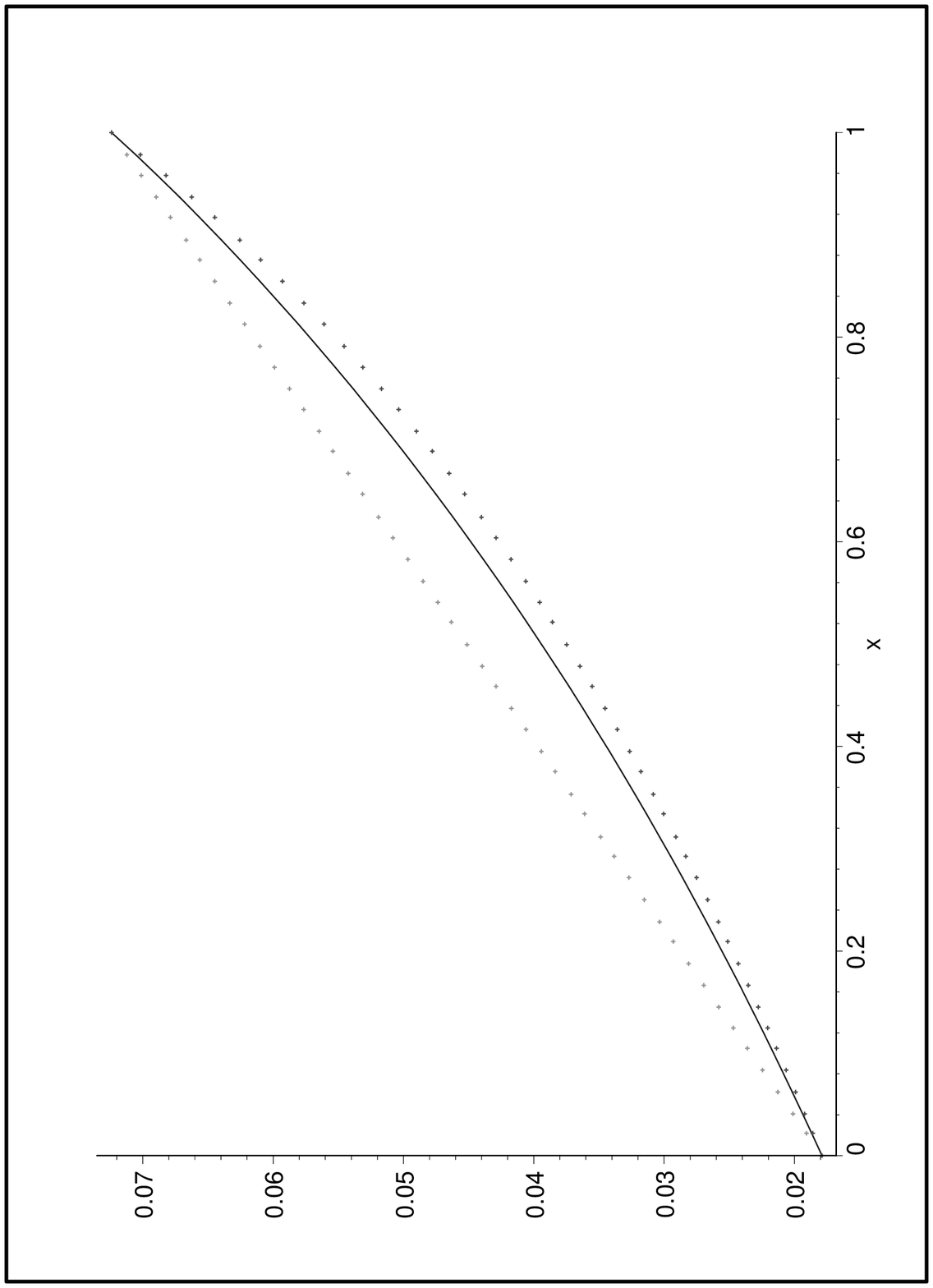, height=8cm}
\end{rotate}


\vspace{8cm}
\begin{center}
\footnotesize{Figure~5: Plot of the surface tension given by (\ref{tauapprox}) (solid curve)
as function of the concentration $c_1$.
The dotted curve corresponds to formula (\ref{taugugg}) and the dotted line to formula (\ref{taudfe}).
}
\end{center}

Finally, notice that when the coupling constants $J_1$ and $J_2$ (or equivalently the surface tensions
$\tau_{1,0}$ and $\tau_{2,0}$) differ appreciably,
we get assuming $J_1 <J_2$, and neglecting in equations ( \ref{tauapprox}) and ( \ref{taugugg}) the terms exponentially small
with respect to the  inverse temperature:
\begin{equation}
  \label{eq:skz}
  \frac { 
\tau_{\text{mixt}, \text{emp}}  
}{
\tau_{1,0} 
}
= 1 -\frac{1}{\beta J_1 }\ln (1-c^*_2)
\end{equation}
to be compared with Szyzkowsky's equation (\ref{eq:sz}).

\section{Proof of Theorem~2}

\setcounter{equation}{0}

\setcounter{theorem}{0}

Let us first give the
proof of relations (\ref{eq:B8})--(\ref{eq:B9}).

Let $\mu ( \Gamma ) = ( a \nu e^{\lambda} )^{- L ( \Gamma )}$ , with $\nu = 2
\nu_d$, $a > 1$, and $\lambda > 0$, then
\begin{equation}
\label{eq:mu}
  \sum_{\Gamma \nsim \Gamma_0} \mu ( \Gamma ) \leq L ( \Gamma_0 ) \sum_{n =
  1}^{\infty} e^{- \lambda n} a^{- n} \leq \frac{e^{- \lambda}}{a - 1} L (
  \Gamma_0 )
\end{equation}
where the sum runs over contours $\Gamma$ incompatible with a given contour
$\Gamma_0$.
\newline
The condition $\nu e^{- ( \alpha - \lambda )} ae^{\frac{1}{a
- 1}} \leq 1$ actually ensures that the convergence condition 
\begin{equation}\label{eq:condconf}
\left| z'_q ( \Gamma_0 )
\right| \leq \left( e^{\mu ( \Gamma_0 )} - 1 \right) \exp \left[ -
\sum_{\Gamma \nsim \Gamma_0} \mu ( \Gamma ) \right]
\end{equation}
 of {\cite{M}} is
fulfilled. We then choose $a = \frac{\sqrt{5} + 3}{2}$ (that minimizes
$ae^{\frac{1}{a - 1}}$)
getting by {\cite{M}} for $\kappa \leq e^{( \alpha - \lambda )}$
the equality (\ref{eq:B8}) and 
\begin{equation}
\label{eq:sal}
   \sum_{X : X ( \Gamma ) \geq 1} | \Phi_q ( X ) | \leq  \mu ( \Gamma )
\end{equation}
The invariance of the
$\Phi_q$ under translations leads to (\ref{eq:BB8}).
On the other hand the last inequality gives
\begin{eqnarray}
  \sum_{X : \text{supp} X \ni x} \left| \Phi_q ( X ) \right| & \leq &
  \sum_{\Gamma : \text{supp} \hspace{0.25em} \Gamma \ni x} \sum_{X : X (
  \Gamma ) \geq 1} \left| \Phi_q ( X ) \right| \\
  & \leq & \sum_{\Gamma : \text{supp} \hspace{0.25em} \Gamma \ni x} \mu (
  \Gamma ) \leq \frac{e^{- \lambda}}{a_0 - 1} \leq \kappa e^{- \alpha} 
  \label{eq:B7}
\end{eqnarray}
by choosing $e^{- \lambda} = \kappa e^{- \alpha}$ and taking into account that
$a_0 - 1 \geq 1$; here the first sum is over all multi-indexes $X$ whose
support contains a given point $x$.
This implies (\ref{eq:BB9}) and also (\ref{eq:B9}) since 
$
 \left| \sigma ( \Lambda \mid \Phi_q ) \right| \leq 
\left| \partial
 \Lambda \right| \sum_{X : \text{supp} \hspace{0.25em} X \ni x} \left| \Phi_q
 ( X ) \right|
$.
Note furthermore that one easily gets the bound (\ref{eq:borneclusterm})
from relations (\ref{eq:mu}) and (\ref{eq:sal})

We shall also gives a bound needed below. Define the diameter of a
contour $\Gamma$ as $\text{diam} \hspace{0.25em} \Gamma = \max_{x, y \in X (
\Gamma )} d ( x, y )$ where $d ( x, y )$ is the distance between the site $x$
and $y$. 
Then
\begin{equation}
  \label{eq:B11} 
\sum_{X : \text{supp} X \ni x \atop
  \text{diam{\hspace{0.25em}}supp} \hspace{0.25em} X \geq A} \left| \Phi_q ( X
  ) \right| \leq \sum_{\Gamma : S ( \Gamma ) \ni x \atop \text{diam}
  \hspace{0.25em} \Gamma \geq A} \mu ( \Gamma ) \leq \sum_{n \geq A}^{\infty}
  e^{- \lambda n} a^{- n} \leq \left( \kappa e^{- \alpha} \right)^A
\end{equation}

We now turn to the proof
of  Theorem~\ref{unicity}. We put $h_q=-p_q$ and 
\begin{equation}
  \label{eq:a}
  a_q=h_q-\min_m h_m
\end{equation}
The boundary condition $q$ is  called stable if $a_q=0$.
Our first step is  to show that if the boundary condition $q$ is stable
then all $q$--contours are stable implying that 
$\Xi _{q}'(\Lambda ;\mu _{1},\mu _{2})$
coincides with
$\Xi _{q}(\Lambda ;\mu _{1},\mu _{2})$.

We notice that 
when  $a_q \leq 1$, then the condition
   \begin{equation}
    \label{eq:stability} \frac{\Xi_m ( \text{Int}_m \hspace{0.25em} \Gamma ;
     \mu_1, \mu_2 )}{\Xi_q ( \text{Int}_m \hspace{0.25em} \Gamma ; \mu_0, \mu_1
     )} \leq e^{2| \partial \text{Int}_m \hspace{0.25em} \Gamma |}
   \end{equation}
   with $m \not= q$ for a $q$--contour $\Gamma$ ensures that this contour is
   stable, provided
   \[ e^{- \alpha} \equiv e^{- \beta J + 5} < \frac{1}{\kappa} \]

 Indeed by (\ref{eq:BB9}) $g_{\max} - g_q$ is bounded by $a_q + 2 \kappa
 e^{- \alpha}$. 
 Since $| \partial \text{Int}_m \hspace{0.25em} \Gamma | \leq |L
 ( \Gamma ) |$ we get taking into account the Peierls estimate
 (\ref{eq:peierls1}):
 \begin{eqnarray*}
   z_q ( \Gamma ) & = & \omega ( \Gamma ) e^{- g_{\max} |S ( \Gamma ) |} e^{(
   g_{\max} - g_q ) |S ( \Gamma ) |} \frac{\Xi_m ( \text{Int}_{\text{mixt}}
   \hspace{0.25em} \Gamma ; \mu_1, \mu_2 )}{\Xi_q ( \text{Int}_{\text{mixt}}
   \hspace{0.25em} \Gamma ; \mu_1, \mu_2 )}\\
   & \leq & e^{- ( \beta J - 5 ) L ( \Gamma )}
 \end{eqnarray*}

For a volume \( \Lambda  \), 
we use \( \text {diam}\, \Lambda =\max _{x,y\in \Lambda }d(x,y) \)
to denote its diameter
The following proposition show that  all
$q$--contours are stable whenever $a_q=0$. 
 \begin{proposition}
   \label{T:stability}Assume $\beta$ is large enough so that $e^{- \beta J + 5}
   = e^{- \alpha} < \frac{1}{( 2 d + 1 ) \kappa}$, then
\begin{description}
\item [(i)]if \( a_{m}\text {diam}\, \Lambda \leq 1 \), and \( a_{q}=0 \),
then \begin{equation}
\label{eq:C1}
\frac{\Xi _{m}(\Lambda ;\mu _{1},\mu _{2})}{\Xi _{q}(\Lambda ;\mu_{1},\mu_{2})}
\leq e^{a_{m}|\Lambda |+2\kappa e^{-\alpha }\left| \partial \Lambda \right| }
\end{equation}

\item [(ii)]if \( a_{q}=0 \), then \begin{equation}
\label{eq:C2}
\frac{\Xi _{m}(\Lambda ;\mu _{1},\mu _{2})}{\Xi _{q}(\Lambda ;\mu_{1},\mu_{2})}
\leq e^{3\kappa e^{-\alpha }\left| \partial \Lambda \right| }
\end{equation}

\item [(iii)]if \( a_{m}\text {diam}\, \Lambda \leq 1 \), then 
\begin{equation}
\label{eq:C3}
\frac{\Xi _{\widetilde{m}}(\Lambda ;\mu _{1},\mu _{2})}{\Xi
  _{m}(\Lambda ;\mu _{1},\mu _{2})}
\leq e^{(1+5\kappa e^{-\alpha })\left| \partial \Lambda \right| }
\end{equation}

\end{description}
  \end{proposition}

The proof is analog to that of Theorem~3.1 in \cite{BI} using our
previous estimates. We give it below for the reader's convenience.
 
We first introduce the notion
of small and large contours. We say that a \( m \)--contour \( \Gamma  \)
is small if \( a_{m} \)diam\( \, \Gamma \leq 1 \); it is large if
\( a_{m} \)diam\( \, \Gamma >1 \). We also define the partition
function \( \Xi _{q}^{\text {small}}(\Lambda ) \) which is obtained
from \( \Xi _{q}^{\prime }(\Lambda ) \) by replacing the sum over
stable contours in (\ref{eq:tpf}) by a sum over small contours. 
If we sum instead, only over contours which are at the same time small
and stable, we denote the resulting partition function 
\( \Xi _{q}^{\prime \text {small}}(\Lambda ) \).
Finally we will use the shorthand notation \( \Xi _{m}(\Lambda ) \)
for \( \Xi _{m}(\Lambda ;\mu _{1},\mu _{2}) \).

We will show the three items of the proposition inductively on diam\( \, \Lambda  \). 

Thus we assume that (i), (ii), and (iii) have already been proved for
all volumes with diam\( \, \Lambda <k \).

\textit{Proof of (i) for} diam\( \, \Lambda =k \)

For any contour \( \Gamma  \) in \( \Lambda  \), and any \( \widetilde{m} \),
we have \( \text {diam}\, \text {Int}_{\widetilde{m}}\, \Gamma \leq k-1 \).
We can use the inductive assumptions (ii) and (iii) that all \( q \)--contours
and all \( m \)--contours are stable. Therefore\begin{equation}
\label{eq:C4}
\frac{\Xi _{q}(\Lambda )}{\Xi _{m}(\Lambda )}
=\frac{\Xi _{q}^{\prime }(\Lambda )}{\Xi _{m}^{\prime }(\Lambda )}
\end{equation}
 Using the convergence of cluster expansion (\ref{eq:BB9}) and
definition (\ref{eq:a}), one immediately gets (i).

\textit{Proof of (ii) for} diam \( \Lambda =k \) 

To control the ratio \( \Xi _{m}(\Lambda )/\Xi _{q}(\Lambda ) \),
we shall rewrite the partition function \( \Xi _{m}(\Lambda ) \)
using relation (\ref{eq:expvap},\ref{eq:expmix}). Consider for a
set of compatible \( m \)--contours in \( \Lambda  \), the family
\( \{\Gamma _{1},\ldots ,\Gamma _{n}\}_{\text {ext}}^{\text {large}} \)
of its mutually external large \( m \)-contours. The others contours
are small \( m \)--contours in 
\( \text {Ext}\equiv \Lambda \setminus \cup _{i}V(\Gamma _{i}) \)
or any \( m \)-contour in \( \text {Ext}\equiv \cup _{i}\text {Ext}\, \Gamma _{i} \).
Therefore\begin{equation}
\Xi _{m}(\Lambda )
=\sum _{\{\Gamma _{1},\ldots ,\Gamma _{n}\}_{\text {ext}}^{\text
    {large}}}
\Xi _{m}^{\text {small}}(\text {Ext})
\prod _{i=1}^{n}\omega (\Gamma _{i})\Xi _{m}(\text {Int}\, \Gamma _{i})
\end{equation}
 Dividing and multiplying by 
\( \prod _{i=1}^{n}e^{g_{\max }|S(\Gamma _{i})|}
\Xi _{q}(\text {Int}\, \Gamma _{i})=\Xi _{q}(\text {Int})\prod _{i=1}^{n}e^{g_{q}|S(\Gamma _{i})|} \),
we get 
\begin{equation}
  \label{eq:C6}
 \frac{\Xi _{m}(\Lambda )}{\Xi _{q}(\Lambda )} 
 =  
\sum _{\{\Gamma _{1},\ldots ,\Gamma _{n}\}_{\text {ext}}^{\text
    {large}}}
\frac{\Xi _{m}^{\text {small}}(\text {Ext})
\Xi _{q}(\text {Int})}{\Xi _{q}(\Lambda )}\prod
_{i=1}^{n}e^{g_{q}|S(\Gamma _{i})|}
\omega (\Gamma _{i})e^{-g_{\max }S(\Gamma _{i})|}
\frac{\Xi _{m}(\text {Int}\, \Gamma _{i})}{\Xi _{q}(\text {Int}\,
  \Gamma _{i})} 
\end{equation}
Note that all \( q \)--contours in \( \Lambda  \) and all small
\( m \)--contours in \( \Lambda  \) are stable by the inductive
assumptions (i) and (iii) respectively. Therefore the various partition
functions in the first factor of the right-hand side of (\ref{eq:C6})
are equal to the corresponding truncated partition functions, which
can be controlled by convergent cluster expansion. We get by (\ref{eq:BB9})
\begin{eqnarray*}
\frac{\Xi _{m}^{\text {small}}(\text {Ext})\Xi _{q}(\text {Int})}{\Xi
  _{q}(\Lambda )}
\prod _{i=1}^{n}e^{g_{q}|S(\Gamma _{i})|} 
& \leq  & e^{-h_{m}^{\text {small}}
\left| \text {Ext}\right| 
+h_{q}\left| \Lambda \setminus \text {Int}\right| }
\prod _{i=1}^{n}e^{g_{q}|S(\Gamma _{i})|}
\\
 &  & \times 
e^{\kappa e^{-\alpha }
(\left| \partial \Lambda \right| 
+\left| \partial \text {Int}\right| 
+\left| \partial \text {Ext}\right| )}
\end{eqnarray*}
where 
\( h_{m}^{\text {small}} \) 
is the free energy obtained from
\( \Xi _{m}^{\text {small}} \). 
Using the facts that 
\( |h_{q}+g_{q}|\leq \kappa e^{-\alpha } \),
\( |S(\Gamma )|\leq |L(\Gamma )| \) 
and bounding 
\( \left| \partial \text {Int}\right| +\left| \partial \text {Ext}\right|  \)
by 
\( \left| \partial \Lambda \right| +2d\sum _{i}L(\Gamma _{i}) \),
we find 
\begin{equation}
\frac{\Xi _{m}^{\text {small}}(\text {Ext})\Xi _{q}(\text {Int})}{\Xi
  _{q}(\Lambda )}
\prod _{i=1}^{n}e^{g_{q}|S(\Gamma _{i})|} 
\leq  e^{-(h_{m}^{\text {small}}-h_{q})\left| \text {Ext}\right| }
e^{\kappa e^{-\alpha }[2\left| \partial \Lambda \right| +(1+2d)\sum _{i}L(\Gamma _{i})]}
\end{equation}

Combining this bound with (\ref{eq:C6}), the Peierls estimates (\ref{eq:peierls1}),
the inductive assumption (ii), and \( \partial \text {Int}\, \Gamma \leq 2dL(\Gamma ) \),
we get:

\begin{eqnarray}
\frac{\Xi ^{m}(\Lambda )}{\Xi ^{q}(\Lambda )} 
& \leq  & e^{2\kappa e^{-\alpha }\left| \partial \Lambda \right| }
\sum _{\left\{ \Gamma _{1},...,\Gamma _{n}\right\} _{\text
    {ext}}^{\text {large}}}
e^{-(h_{m}^{\text {small}}-h_{q})\left| \text {Ext}\right| }
\prod _{i=1}^{n}e^{-\beta J|L(\Gamma _{i})|}e^{[(8d+1)\kappa
  e^{-\alpha }]L(\Gamma _{i})}
\nonumber \\
 & \leq  & e^{2\kappa e^{-\alpha }\left| \partial \Lambda \right| }
\sum _{\left\{ \Gamma _{1},...,\Gamma _{n}\right\} _{\text
    {ext}}^{\text {large}}}
e^{-(h_{m}^{\text {small}}-h_{q})\left| \text {Ext}\right| }
\prod _{i=1}^{n}e^{-(\alpha +1)L(\Gamma _{i})}\label{eq:C9} 
\end{eqnarray}
where for the last inequality we used the hypothesis 
\( (2d+1)\kappa e^{-\alpha }\leq 1 \)
and \( \alpha =\beta J-5 \).

At this point we need a technical lemma proved in \cite{Z} (see the
proof below)

\begin{lemma}

\label{L:Z}

Consider the partition function\begin{equation}
\label{eq:C10}
\widetilde{\mathcal{Z}}(\Lambda )
=\sum _{\left\{ \Gamma _{1},...,\Gamma _{n}\right\} _{\text {comp}}}
\prod _{i=1}^{n}\widetilde{z}(\Gamma _{i})e^{L(\Gamma _{i})}
\end{equation}
 of a gas of contours with activities 
\[
\widetilde{z}(\Gamma )e^{|L(\Gamma )|}\leq e^{-\widetilde{\alpha }L(\Gamma )}e^{L(\Gamma )}\]
 Let \( -\widetilde{s}=-\lim _{\Lambda \uparrow \mathbb {Z}^{d}}\left( 1/|\Lambda |\right)  \)
\( \ln \widetilde{\mathcal{Z}}(\Lambda ) \) be the corresponding
free energy. Then\begin{equation}
\label{eq:C11}
\sum _{\left\{ \Gamma _{1},...,\Gamma _{n}\right\} _{\text {ext}}}
e^{-a\left| \text {Ext}\right| }
\prod _{i=1}^{n}\widetilde{K}(\Gamma _{i})
\leq e^{\kappa e^{-\widetilde{\alpha }+1}|\partial \Lambda |}
\end{equation}
where the sum is over contours in \( \Lambda  \) provided \begin{equation}
\label{eq:C12}
a\geq \widetilde{s}
\end{equation}

\begin{equation}
\label{eq:C:13}
\kappa e^{-\widetilde{\alpha }+1}\leq \frac{1}{2d+1}
\end{equation}
\end{lemma}

To apply this lemma to (\ref{eq:C9}) we put 
\begin{eqnarray}
\widetilde{\alpha } 
& = & \alpha +1\label{eq:C14} \\
a & = & h_{m}^{\text {small}}-h_{p}=a_{m}+h_{m}^{\text {small}}-h_{m}
\end{eqnarray}
 and\begin{equation}
\label{eq:C16}
\widetilde{z}(S)
=\left\{ \begin{array}{ll}
e^{-\widetilde{\alpha }L(\Gamma )} 
& \text {if}\quad \Gamma \quad \text {is}\, \text {large}\\
0 & \text {if}\quad \Gamma \quad \text {is}\, \text {small}
\end{array}\right. 
\end{equation}

For \( \kappa e^{-\widetilde{\alpha }+1}<1 \), the Mayer expansion
for \( \ln \widetilde{\mathcal{Z}}(\Lambda ) \) is convergent. Using
the fact that it only contains large contours (\( \text {diam}\, \Gamma \geq 1/a_{m} \)
for each contour contributing to \( \ln \widetilde{\mathcal{Z}}(\Lambda ) \))
one has (see (\ref{eq:B11})):\begin{equation}
\label{eq:C18}
\widetilde{s}\leq \left( \kappa e^{-\alpha }\right) ^{\frac{1}{a_{m}}}
\end{equation}
 Moreover the difference \( h_{m}^{\text {small}}- \) \( h_{m} \)
is the free energy of a gas of large contours with again \( \text {diam}\, \Gamma \geq 1/a_{m} \)
and thus again for \( \kappa e^{-\widetilde{\alpha }-1}<1 \) shares
the same upper bound 
\begin{equation}
\label{eq:C19}
\left| h_{m}^{\text {small}}-h_{m}\right| \leq \left( \kappa e^{-\alpha }\right) ^{\frac{1}{a_{m}}}
\end{equation}
Hence 
\begin{equation}
\label{eq:C20}
a-\widetilde{s}\geq a_{m}-2\left( \kappa e^{-\alpha }\right) ^{\frac{1}{a_{m}}}
\end{equation}
 Therefore the assumption \( a-\widetilde{s}\geq 0 \) will be fulfilled
if \( a_{m}\geq 2\left( \kappa e^{-\alpha }\right) ^{\frac{1}{a_{m}}} \)
i.e. if \( (a_{m}/2)^{a_{m}}\geq \kappa e^{-\alpha } \). This is
actually true because the function \( (a/2)^{a} \) has a minimum
at \( a=2/e \) for which it takes the value \( e^{-2/e}\simeq 0.47 \)
that is greater than the upper bound \( 1/(2d+1) \) required for
\( \kappa e^{-\alpha } \).

Applying the lemma (with the value (\ref{eq:C14})) to(\ref{eq:C9})
immediately gives (\ref{eq:C2}).

\textit{Proof of (iii) for} diam \( \Lambda =k \)

The inequality (\ref{eq:C3}) follows immediately from (\ref{eq:C1}),
(\ref{eq:C2}) , and the fact that
\begin{equation}
a_{m}|\Lambda |
\leq a_{m}\text {diam}\, \Lambda \left| \partial \Lambda \right| 
\leq \left| \partial \Lambda \right| 
\end{equation}

\textit{Proof of Lemma \ref{L:Z}} 
For \( \kappa e^{-\widetilde{a}+1}<1 \),
the partition function \( \widetilde{\mathcal{Z}}(\Lambda ) \) can
be controlled by convergent cluster expansion. In particular for the
interior Int\( =\cup _{i}\text {Int}\, \Gamma _{i} \) of a set of
external contours \( \left\{ \Gamma _{1},...,\Gamma _{n}\right\} _{\text {ext}} \)
we have the estimate
\begin{equation}
\widetilde{\mathcal{Z}}(\text {Int})
e^{-\widetilde{s}|\text {Int}|}\geq 
e^{-\kappa e^{-\widetilde{\alpha }+1}
\left| \partial \text {Int}\right| }
\geq \prod _{i=1}^{n}e^{-2d\kappa e^{-\widetilde{\alpha }+1}|L(\Gamma _{i})|}
\end{equation}
 Here the first inequality stems from (\ref{eq:B9}) and the last
from the hypothesis (\ref{eq:C:13}). Therefore 
\begin{eqnarray*}
 \sum _{\left\{ \Gamma _{1},...,\Gamma _{n}\right\} _{\text
     {ext}}}
e^{-a\left| {Ext}\right| }\prod _{i=1}^{n}\widetilde{z}(\Gamma _{i})
&\leq &
\sum _{\left\{ \Gamma _{1},...,\Gamma _{n}\right\} _{\text
     {ext}}}
e^{-a\left| \text {Ext}\right| }\prod _{i=1}^{n}\widetilde{z}(\Gamma
     _{i})
e^{2d\kappa e^{-\widetilde{\alpha }+1}L(\Gamma _{i})}
\widetilde{\mathcal{Z}}(\text {Int})e^{\widetilde{s}|\text {Int}|}
\\
 &\leq  & e^{-\widetilde{s}|\Lambda |}
\sum _{\left\{ \Gamma _{1},...,\Gamma _{n}\right\} _{\text {ext}}}
\prod _{i=1}^{n}\widetilde{z}(\Gamma _{i})e^{(2d\kappa
     e^{-\widetilde{\alpha }+1}
+\widetilde{s})L(\Gamma _{i})}\widetilde{\mathcal{Z}}(\text {Int})\\
 &  \leq & e^{-\widetilde{s}|\Lambda |}
\widetilde{\mathcal{Z}}(\text {Int})
\end{eqnarray*}
where in the last inequality, we used 
\( \widetilde{s}\leq \kappa e^{-\widetilde{\alpha }+1} \).
This gives (\ref{eq:C11}) using again the estimates (\ref{eq:B9}).
$\Box$
 
This ends the proof of  Proposition~\ref{T:stability}.

We now come back to the proof of Theorem~\ref{unicity}.
We put $\mu_1(t)=\mu_1 +t$, $\mu_2(t) =\mu_2 +t$ with
$e^{\mu _{1}}+e^{\mu _{2}}=1$.
Then,
by definitions (\ref{eq:tpf}),(\ref{eq:thlm}),   and (\ref{eq:a}),
we have 
\begin{equation}
\label{eq:D2}
a_{\text{emp} }- a_{\text{mixt} }
=p_{\text{mixt} }-p_{\text{emp} }
=t+\ln (e^{\mu _{1}}+e^{\mu _{2}})
+\lim _{\Lambda \uparrow \mathbb {Z}^{d}}\frac{1}{|\Lambda |}
\left[ \ln \mathcal{Z}_{\text{mixt} }(\Lambda )-\ln \mathcal{Z}_{\text{emp} }(\Lambda )\right] 
\end{equation}
where
\begin{equation}
\label{eq:tpfcm} \mathcal{Z}_q ( \Lambda ) = e^{- g_q | \Lambda |} \Xi'_q (
  \Lambda ) = \sum_{\left\{ \Gamma_1, ..., \Gamma_n \right\}_{\text{comp}}}
  \prod_{i = 1}^n z_q' ( \Gamma_i )
 \end{equation}
 The function  \(t+\ln (e^{\mu _{1}}+e^{\mu _{2}})=t\)  is obviously increasing, 
negative for \(t<0 \),
positive for \(t>0 \), and it intersects
the horizontal coordinate axis only at one point \(t=0 \).
The difference \( a_{\text{mixt} }-a_{\text{emp} } \) will satisfy the same
properties with the intersecting point slightly changed) provided
\begin{equation}
\label{eq:resultat}
\frac{1}{|\Lambda |}\left| \frac{\partial }{\partial t}
\ln \mathcal{Z}_{\text{emp}}(\Lambda )-\frac{\partial }{\partial t}
\ln \mathcal{Z}_{\text{mixt} }(\Lambda )\right| <1
\end{equation}
uniformly in \( \Lambda  \).

Let us first give an upper bound on the derivative \( \frac{\partial }{\partial t}z'_{q}(\Gamma ) \)
of the truncated activity. By virtue of relations (\ref{eq:stable}),(\ref{eq:activity})
one gets for every stable \( q \)--contour \( \Gamma  \), that either
\( \frac{\partial }{\partial t }z'_{q}(\Gamma )=0 \), or: 
\begin{eqnarray}
\left| \frac{\partial }{\partial t}z'_{q}(\Gamma )\right|  
& = & \Bigg |\frac{\partial }{\partial t}
\ln \omega (\Gamma )-
\frac{\partial }{\partial t}g_{q}(\Gamma )+\frac{\partial
}{\partial t}
\ln \frac{\Xi _{m}(\text {Int}_{m}\, \Gamma )}{\Xi _{q}(\text
  {Int}_{m}\, \Gamma )}
\Bigg |z'_{q}(\Gamma )\nonumber \\
 & \leq  & (1+|\text {Int}_{m}\, \Gamma |)z'_{q}(\Gamma )
\leq |V(\Gamma )|e^{-\alpha |L(\Gamma )|}\label{eq:D3} 
\end{eqnarray}
 On the other hand, 
\begin{equation}
\label{eq:D4}
\left| \frac{1}{|\Lambda |}\frac{\partial }{\partial t}
\ln \mathcal{Z}_{q}(\Lambda )\right| 
\leq 
\sum _{\Gamma :\text {supp}\, \Gamma \ni x}\left| 
\frac{\partial }{\partial t}z'_{q}(\Gamma )\right| 
\times \left| 
\frac{\mathcal{Z}_{q}(\Lambda \setminus \left\{ \Gamma \right\}
  )}{\mathcal{Z}_{q}(\Lambda )}
\right| 
\end{equation}
 Here the sums are over contours \( \Gamma  \) containing a given
point \( x \) and 
\begin{equation}
\mathcal{Z}_{q}(\Lambda \setminus \left\{ \Gamma \right\} )
=\sideset {}{^{*}}\sum _{\left\{ \Gamma _{1},...,\Gamma _{n}\right\} _{\text {comp}}}z'_{q}(\Gamma )
\end{equation}
 where the sum goes over all families 
\( \left\{ \Gamma _{1},...,\Gamma _{n}\right\} _{\text {comp}} \)
compatible with \( \Gamma  \). 
This sum can be bounded using the
cluster expansion. Indeed, denoting \( \overline{S} \) the set of
sites at distance less or equal to \( 1 \) from the support of \( \Gamma  \),
we get by (\ref{eq:B8}),(\ref{eq:B9}): 
\begin{eqnarray}
\frac{\mathcal{Z}_{q}(\Lambda \setminus \left\{ \Gamma \right\}
  )}{\mathcal{Z}_{q}(\Lambda )} 
& = & \exp \left\{ -\sum _{X:\text {supp}X\cap \overline{S}=\emptyset
  }
\Phi _{q}(X)\right\} 
\leq \exp \left\{ |\overline{S}|
\sum _{X:\text {supp}\, X\ni x}\left| \Phi _{q}(X)\right| \right\} 
\nonumber \\
 & \leq  & \exp \left\{ (2d+1)|S(\Gamma )|\kappa e^{-\alpha }\right\} 
\leq \exp \left\{ (2d+1)L(\Gamma )\kappa e^{-\alpha }\right\} \label{EQ:D7} 
\end{eqnarray}

Inserting this bound in (\ref{eq:D4}) and taking into account the
inequality (\ref{eq:D3}) and the estimate 
\[
|V(\Gamma )|\leq |S(\Gamma )|\text {diam}\, \Gamma \leq |S(\Gamma
)|^{2}
\leq e^{|S(\Gamma )|}\leq e^{L(\Gamma )}\]
gives
\[
\left| 
\frac{1}{|\Lambda |}
\frac{\partial }{\partial t}
\ln \mathcal{Z}_{q}(\Lambda )
\right|  
 \leq   
\sum _{\Gamma :\text {supp}\, \Gamma \ni x}
e^{-\alpha L(\Gamma
  )+(2d+2)L(\Gamma )Ke^{-\alpha }}
  \leq   \sum _{n\geq 1}\nu ^{n}e^{-\alpha n}e^{2(d+1)\kappa e^{-\alpha }n}
\]
Using that
\( \kappa \simeq 4.9\nu  \), 
we
get that for 
\( 2(d+1)\kappa e^{-^{\alpha }}\leq 1 \) 
this last sum
is less than \( 1/2 \) .
This implies (\ref{eq:resultat}) 
ending
the proof of Theorem~\ref{unicity}. \ \rule{0.5em}{0.5em}

\section{Proof of Theorem~3}

\setcounter{equation}{0}

\setcounter{theorem}{0}

Let us first explain the idea of the proof.

In dimension $d = 3$, the first excitation of a flat
interface is obtained by replacing an empty site of the interface by an
occupied one (denote $I^{\text{up}}_{\text{el}}$ such an interface) or by
replacing an occupied site of the interface by an empty one (denote
$I^{\text{down}}_{\text{el}}$ such an interface). In the first case, we
have
\begin{eqnarray*}
  &  & \sum_{I : \text{supp} \hspace{0.25em} I \subset \overline{V} \atop
  \text{supp} \hspace{0.25em} I = \text{supp} \hspace{0.25em}
  I^{\text{up}}_{\text{el}}} \omega ( I ) e^{- p ( \mu_1^{\ast},
  \mu_2^{\ast} ) |S ( I ) |}\\
  &  & = \sum_{n = 0}^{N - 1} \binom{N - 1}{n} \frac{e^{( \mu_1^{\ast}
  - \beta J_1 ) n} e^{- ( \mu_2^{\ast} - \beta J_2 ) ( N - 1 - n )}}{e^{p (
  \mu_1^{\ast}, \mu_2^{\ast} ) N}} ( e^{\mu_1^{\ast} - 5 \beta
  J_1} + e^{\mu_2^{\ast} - 5 \beta J_2} )\\
  &  & = ( c^*_1 e^{- \beta J_1} + c^*_2 e^{- \beta J_2} )^N \frac{c^*_1 e^{- 5
  \beta J_1} + c^*_2 e^{- 5 \beta J_2}}{c^*_1 e^{- \beta J_1} + c^*_2 e^{- \beta
  J_2}}
\end{eqnarray*}

In the second case we have
\begin{eqnarray*}
  &  & \sum_{I : \text{supp} \hspace{0.25em} I \subset \overline{V} \atop
  \text{supp} \hspace{0.25em} I = \text{supp} \hspace{0.25em}
  I^{\text{down}}_{\text{el}}} \omega ( I ) e^{- p ( \mu_1^{\ast},
  \mu_2^{\ast} ) |S ( I ) |}\\
  &  & 
= \sum_{n = 0}^{N - 4} \frac{\binom{N - 4}{n} e^{( \mu_1^{\ast}
  - \beta J_1 ) n} 
e^{- ( \mu_2^{\ast} - \beta J_2 ) ( N - 4 - n )}}{e^{p (
  \mu_1^{\ast}, \mu_2^{\ast} ) N}} 
\sum_{m = 1}^4 \binom{4}{m} e^{2 m (
  \mu_1^{\ast} - \beta J_1 )} e^{- ( 8 - 2 m ) ( \mu_2^{\ast} - \beta J_2 )}
\\
  &  & \\
  &  & = ( c^*_1 e^{- \beta J_1} + c^*_2 e^{- \beta J_2} )^N \frac{( c^*_1 e^{- 2
  \beta J_1} + c^*_2 e^{- 2 \beta J_2} )^4}{( c^*_1 e^{- \beta J_1} + c^*_2 e^{-
  \beta J_2} )^4}
\end{eqnarray*}

In dimension $d = 2$, the first excitation of a flat
interface is obtained by splitting it into a left part and a right part and
then shift the right part by a height $1$ or $- 1$ (denote $I_{\text{el}}$
such an interface). Then, we have
\begin{eqnarray*}
  &  & 
\sum_{I : \text{supp} \hspace{0.25em} I \subset \overline{V} \atop
  \text{supp} \hspace{0.25em} I = \text{supp} \hspace{0.25em} I_{\text{el}}}
  \omega ( I ) e^{- p ( \mu_1^{\ast}, \mu_2^{\ast} ) |S ( I ) |}\\
  &  & = 
\sum^{N - 1}_{n = 0} 
\binom{N - m}{n} 
\frac{e^{( \mu_1^{\ast} - \beta J_1 ) n} 
e^{- ( \mu_2^{\ast} - \beta J_2 ) ( N - m - 1 )}
}{e^{p (
  \mu_1^{\ast}, \mu_2^{\ast} ) N}} ( e^{\mu_1^{\ast} - 2 \beta J_1} +
  e^{\mu_2^{\ast} - 2 \beta J_2} )\\
  &  & = ( c^*_1 e^{- \beta J_1} + c^*_2 e^{- J_2} )^N 
\frac{c^*_1 e^{- 2 \beta
  J_1} 
+ c^*_2 e^{- 2 \beta J_2}}{c^*_1 e^{- \beta J_1} + c^*_2 e^{- \beta J_2}}
\end{eqnarray*}
Note, that in fact such modified interfaces are not allowed with
the boundary condition $\chi^{\text{mixt},\text{emp}}$.  
However, one can lightly modify this boundary condition to allow such interface,
leaving the resulting surface tension unchanged.

To study the difference 
$F' (  \overline{V} ) -
F_{\text{flat}} ( \overline{V} )$,
we shall express the quantity
\[ 
( c^*_1 e^{- \beta
   J_1} + c^*_2 e^{- \beta J_2} )^{- N} \sum_{I : \text{supp} \hspace{0.25em} I
   \subset \overline{V}} \omega ( I ) e^{- p ( \mu_1^{\ast},
   \mu_2^{\ast} ) |S ( I ) |} 
\]
as the partition function of a gas of
excitations to be called walls or more generally aggregates with small activities at low enough
temperature. 

This will allow us to exponentiate this quantity and to obtained
that the difference 
$( F' ( \overline{V} ) - F_{\text{flat}}' ( \overline{V} ) ) / N$
can be
expressed as a convergent series (up to a boundary term) whose leading terms
are :
\[ 
-\frac{1}{\beta}
\frac{  c^*_1 e^{- 5 \beta J_1} + c^*_2 e^{- 5 \beta J_2}
}{
     c^*_1 e^{- \beta J_1} + c^*_2 e^{- \beta J_2}
} 
-\frac{1}{\beta}
\frac{ ( c^*_1 e^{- 2 \beta J_1} + c^*_2 e^{- 2 \beta J_2} )^4
}{
(   c^*_1 e^{- \beta J_1} + c^*_2 e^{- \beta J_2}  )^4
}
\]
in
three dimensions and
\[ 
-\frac{2}{\beta} \frac{c^*_1 e^{- 2 \beta J_1} + c^*_2 e^{- 2 \beta
   J_2}}{c^*_1 e^{- \beta J_1} + c^*_2 e^{- \beta J_2}} 
\]
in two dimensions.

\subsection{Decorated  interfaces}
The first step is to express the quantity $A ( I )$
defined by (\ref{eq:aa}) in a form suitable for
our purpose.
We write for each support of the  mixt or $\text{emp}$ cluster $C$ in $A(I)$:
\begin{eqnarray*}
 e^{- \tilde{\Phi}_{\text{mixt}} ( C ) 
\frac{|C \cap D|}{|C|}}
& = & 
1+
( e^{- \tilde{\Phi}_{\text{mixt}} ( C ) 
\frac{|C \cap D|}{|C|}} - 1 )
\equiv 1+ \tilde{\psi}_{\text{mixt}} ( C) 
\\
 e^{- \tilde{\Phi}_{\text{emp}} ( C ) 
\frac{|C \cap U|}{|C|}}
& = & 
1+
( e^{- \tilde{\Phi}_{\text{emp}} ( C ) 
\frac{|C \cap U|}{|C|}} - 1 )
\equiv 1+ \tilde{\psi}_{\text{emp}} ( C) 
 \end{eqnarray*}
Define for a connected family $A$ of support of clusters:
\[
  \tilde{\psi}_{\text{mixt}} ( A ) 
 =  
\prod_{C \in A}  \tilde{\psi}_{\text{mixt}} ( C), \quad 
  \tilde{\psi}_{ \text{emp} } ( A ) 
 =  
\prod_{C \in A}  \tilde{\psi}_{ \text{emp} } ( A ) 
\]
Then
\begin{eqnarray*}
 A(I)
& = & 
\prod_{C \cap S ( I ) \not= \emptyset} 
e^{- \tilde{\Phi}_{\text{mixt}} ( C ) \frac{|C \cap D|}{|C|}}
  \prod_{C' \cap S ( I ) \not= \emptyset} 
e^{- \tilde{\Phi}_{\text{emp}} (  C' ) \frac{|C' \cap U|}{|C|}}
\\
  & = &
\sum_{\{ A_1, ..., A_n \}_{\text{comp}} : A_i \cap S ( I ) \not= \emptyset}
   \prod_{i = 1}^n \tilde{\psi}_{\text{mixt}} ( A_i ) 
\sum_{\{  A_1, ..., A_m \}_{\text{comp}} : A_i \cap S ( I ) \not= \emptyset}
  \prod_{j = 1}^m \tilde{\psi}_{\text{emp}} ( A_j )
\end{eqnarray*}
where the sums are over compatible families of connected sets $A_i$ of support of clusters
touching the interface.
As it was done for multi-indexes, it is convenient to
sum  all  $A$  with the same support say  $\mathcal{D}$ to be called decoration. 
We define the
weight
\begin{eqnarray}
  \psi_{\text{mixt}} (\mathcal{D} ) & = & 
\sum_{A : \text{supp}\, A = \mathcal{D}} \tilde{\psi}_{\text{mixt}} ( A )
=\sum_{\left\{ C_1, ..., C_n \right\} : \cup
  C_i = \mathcal{D}}\prod_{i=1}^{n} \tilde{\psi}_{\text{mixt}} ( C_i) 
\\
  \psi_{\text{emp}}(\mathcal{D} ) & = &
\sum_{A : \text{supp}\, A = \mathcal{D}}\tilde{\psi}_{\text{emp}} ( A ) 
= \sum_{\left\{ C_1, ..., C_n \right\} : \cup
  C_i = \mathcal{D}}\prod_{i=1}^{n} \tilde{\psi}_{\text{emp}} ( C_i ) 
\end{eqnarray}
This leads to
\begin{eqnarray}
  e^{- \beta F' ( \overline{V} )}
& = & \sum_{I : \text{supp} \hspace{0.25em} I \subset \overline{V}}
\omega ( I ) e^{-  p ( \mu_1^{\ast},  \mu_2^{\ast} ) |S ( I ) |}  
\sum_{\{  \mathcal{D}_1, ...,  \mathcal{D}_n \}_{\text{comp}} 
: \atop  \mathcal{D}_i \cap S ( I ) \not= \emptyset,  \mathcal{D}_i \cap D \not= \emptyset}
  \prod_{i = 1}^n \psi_{\text{mixt}} (  \mathcal{D}_i )
\nonumber
\\
  &  & 
\times \sum_{\{  \mathcal{D}_1, ...,  \mathcal{D}_m \}_{\text{comp}} 
: \atop  \mathcal{D}_i \cap S ( I ) \not= \emptyset,  \mathcal{D}_i 
\cap U \not= \emptyset} 
\prod_{j = 1}^m  \psi_{\text{emp}} (  \mathcal{D}_j ) 
\label{fprime}
\end{eqnarray}
where the sums are over compatible families of decorations touching the interface. 

We define  a decorated interface as a triplet 
$I^{\text{de}}=\{I, \mathfrak{D}_{\text{mixt}}, \mathfrak{D}_{\text{emp}}\}$,
where $I$ is an interface, $ \mathfrak{D}_{\text{mixt}}$ is a collection of 
mixt--decorations touching the interface and $ \mathfrak{D}_{\text{emp}}$
is a collection of $\text{emp}$--decorations touching the interface.

The weights of decorations may be controlled
with the inequalities 
$|e^{- \tilde{\Phi}_{\text{mixt}} ( C )
\frac{|C \cap
D|}{|C|}} \newline
- 1| \leq ( e - 1 ) | \tilde{\Phi}_{\text{mixt}} ( C ) |$ and $|e^{-
\tilde{\Phi}_{\text{emp}} (  \mathcal{D}) \frac{|C \cap U|}{|C|}} - 1| \leq ( e - 1 ) |
\tilde{\Phi}_{\text{emp}} (  \mathcal{D}) |$. 
Together with (\ref{eq:borneclusterm}),
this implies the bounds (see \cite{DLR}):
\begin{eqnarray}
  | \psi_{\text{mixt}} (  \mathcal{D}) | & \leq & ( 8 e ( e - 1 ) \kappa e^{- \alpha}
  )^{L (  \mathcal{D} )}  \label{bornecl1}\\
  | \psi_{\text{emp}} (  \mathcal{D}) | & \leq & ( 8 e ( e - 1 ) \kappa e^{- \alpha}
  )^{L (  \mathcal{D})}  \label{bornecl2}
\end{eqnarray}

\subsection{Walls and aggregates}
We now introduce the notion of walls and aggregates by the following definitions.

Consider a decorated  interface 
$I^{\text{de}}=\{I, \mathfrak{D}_{\text{mixt}},\mathfrak{D}_{\text{emp}}\}$. 
Let $\Pi_0$ denote the horizontal hyper-plane 
$x_d =0$ and let $\pi$ denote the
projection parallel to the vertical axis on this hyper-plane: 
$\pi ( x_1, ..., x_d ) = ( x_1, ..., x_{d - 1} )$, $\pi ( A)=\cup_{x\in A} \pi ( x)$.
 A pair 
$\left\{ s_x,s_y \right\}$ of the  interface $I $ is called correct if
\begin{enumerate}
  \item $\pi ( x ) = \pi ( y )$ (the bond $xy$ is vertical).
  
  \item there is no other pair pairs $\left\{ s_{x'}, s_{y'} \right\}$ of the
  interface such that $\pi ( x' ) = \pi ( x )$.
  
  \item there are no decorations  such that 
$\pi ( \mathcal{D} ) \supset \pi ( x  )$.
\end{enumerate}
The connected components of the set of correct pairs are
called \textit{ceilings} and denoted $\mathcal{C}$. 
The connected components of the set
of non correct pairs are called \textit{walls} (see Fig.~5). 

For a
wall $W$, we use $S_{\alpha} ( W )$ to denote the set of sites for which 
$s_x= \alpha$. 
The set $\text{supp} \hspace{0.25em} W = S_0 ( W ) \cup S_1 ( W )
\cup S_2 ( W )$ is called support of the  wall $W$. 
As for the contours and the interfaces, we will
also use $S ( W ) = S_1 ( W ) \cup S_2 ( W )$ to denote the set of occupied
sites of the wall, $L_1 ( W )$ (respectively $L_2 ( W )$) to denote the number
of nearest neighbour pairs $\langle x, y \rangle$ such that such that $s_x =
1$ and $s_y = 0$ (respectively 
$s_x = 2$ and $s_y = 0$) 
and $L ( W ) = L_1 ( W
) + L_2 ( W )$.

Denote further $\mathcal{W}(I)$ 
the set of  walls of the interface $I$.

The  union 
$\cup_{W\in \mathcal{W}(I)}\text{supp} \hspace{0.25em} W 
\cup_{\mathcal{D}\in I^{\text{de}} } \mathcal{D} $
split into maximal connected components called aggregates.
Namely, an aggregate $w$ of a decorated interface  is a family 
$$w= \{ W_1, ..., W_n ; \mathcal{D}_1, ..., \mathcal{D}_m \}$$
 such that the set
$\cup_{i=1}^n\text{supp} \hspace{0.25em} W_i \cup_{j=1 }^m \mathcal{D}_j $ is
connected.

A collection of aggregates $\{w_1,\dots,w_n\}$ is called \textit{admissible} 
if  there exists
a decorated interface $I^{\text{de}}$ such that  $w_1,\dots,w_n$ are the aggregates 
of $I^{\text{de}}$. 

Here comes a difference for definitions between dimension $2$ and dimension $3$.

In $3$--dimensions an aggregate
is called a {\it standard aggregate}, (or aggregate in standard position), if there exits 
a decorated interface  $I^{\text{de}}$
such that $w$ is the unique aggregate of  $I^{\text{de}}$.
To any aggregate $w$, we can associate a unique standard aggregate which is just a translate
of $w$.
In this way, one can associate to any admissible collection of aggregates a unique collection
of standard aggregates. Such collections are called admissible collections
of standard aggregates.
To  an admissible collection  of standard aggregates we can associate 
in a unique way an admissible collection of aggregates.

\vspace{1cm}
\begin{center}
\setlength{\unitlength}{20pt}
\ifx\plotpoint\undefined\newsavebox{\plotpoint}\fi
\begin{picture}(15,9)(0,0)
\font\gnuplot=cmr10 at 10pt
\gnuplot

\put(6,3){$\mathbf{1}$}

\put(5,4){$\mathbf{1}$}
\put(6,4){$\mathbf{0}$}
\put(7,4){$\mathbf{2}$}

\put(1,5){$\mathbf{2}$}
\put(2,5){$\mathbf{1}$}

\put(5,5){$\mathbf{1}$}
\put(6,5){$\mathbf{0}$}
\put(7,5){$\mathbf{2}$}

\put(1,6){$\mathbf{0}$}
\put(2,6){$\mathbf{0}$}
\put(3,6){$\mathbf{1}$}

\put(5,6){$\mathbf{2}$}
\put(6,6){$\mathbf{0}$}
\put(7,6){$\mathbf{1}$}

\put(11,6){$\mathbf{1}$}
\put(12,6){$\mathbf{0}$}

\put(1,7){$\mathbf{0}$}
\put(2,7){$\mathbf{0}$}
\put(3,7){$\mathbf{1}$}

\put(5,7){$\mathbf{2}$}
\put(6,7){$\mathbf{0}$}
\put(7,7){$\mathbf{0}$}

\put(11,7){$\mathbf{0}$}

\put(0,8){$\mathbf{0}$}
\put(1,8){$\mathbf{2}$}
\put(2,8){$\mathbf{1}$}
\put(3,8){$\mathbf{1}$}

\put(5,8){$\mathbf{2}$}
\put(6,8){$\mathbf{0}$}

\put(1,9){$\mathbf{0}$}
\put(2,9){$\mathbf{0}$}
\put(3,9){$\mathbf{0}$}

\put(5,9){$\mathbf{0}$}

\drawline(0.7,5.7)(2.7,5.7)
\drawline(2.7,5.7)(2.7,7.7)
\drawline(2.7,7.7)(0.7,7.7)

\drawline(0.7,7.7)(0.7,8.7)
\drawline(0.7,8.7)(3.7,8.7)
\drawline(4.7,8.7)(5.7,8.7)

\drawline(5.7,8.7)(5.7,3.7)
\drawline(5.7,3.7)(6.7,3.7)

\drawline(6.7,3.7)(6.7,6.7)

\drawline(6.7,6.7)(7.7,6.7)
\drawline(10.7,6.7)(11.7,6.7)

\drawline(11.7,6.7)(11.7,5.7)

\end{picture}

\end{center}

\vspace{-2cm}
\begin{center}
\begin{footnotesize}
Figure~6: The walls corresponding to the interface of Figure~4.
\end{footnotesize}
\end{center}
\subsection{Expansions}

With these definitions and notations, 
one gets
from (\ref{fprime}),
 the following expansion
\begin{equation}
\label{wallpf}
  e^{- \beta F' ( \overline{V} )}
= ( c^*_1 e^{- \beta J_1} + c^*_2 e^{- \beta
  J_2} )^N \sum_{\{ w_1, \ldots, w_n \}_{\text{adm}}} \prod_{i = 1}^n z ( w_i
  )
\end{equation}
where the sum is over admissible collections of standard aggregates and the
activities of aggregates are given by
\begin{equation}
\label{act}
z ( w )
=
\prod_{W \in w}
\frac{e^{-
   \beta J_1 L_1 ( W ) - \beta J_2 L_2 ( W ) + \mu_1^{\ast} |S_1 ( W ) | +
   \mu_2^{\ast} |S_2 ( W ) |}
}{e^{p ( \mu_1^{\ast}, \mu_2^{\ast} ) |S (
   W ) |}
( c^*_1 e^{- \beta J_1} + c^*_2 e^{- \beta J_2} )^{| \pi ( W ) |}
}
   \prod_{\mathcal{D} \in w} \psi ( \mathcal{D} )
\end{equation}
Here $\psi ( \mathcal{D}) = \psi_{\text{mixt}} ( \mathcal{D})$ for the
mixt--decorations and $\psi ( \mathcal{D} ) = \psi_{\text{emp}} ( \mathcal{D} )$ for
the $\text{emp}$-decorations.

In two dimensions we proceed differently.
For an aggregate $w$, we consider the ceiling components to the left and to the right of $w$ 
respectively denoted  
$\mathcal{C}^{(L)}$ 
and  
$\mathcal{C}^{(R)}$.
Let 
$h^{(L)}$ 
(respectively 
$h^{(R)}$) be the second coordinate of any empty site of
   $\mathcal{C}^{(L)}$ (respectively $\mathcal{C}^{(R)}$).
We define the position 
$p(w)$ of th aggregate $w$ as 
$p(w)=h^{(L)}$
and the height of the aggregate $w$ as $
h(w)=h^{(R)}-h^{(L)}$.
An aggregate $w$ is now called standard aggregate (or aggregate in standard position) if $p(w)=0$.
One obviously can associate to any aggregate an aggregate in standard position and to any
admissible collection of aggregates a unique collection of standard aggregates. 
As before such collections  are called admissible collections of standard aggregates
and to  an admissible collection of standard aggregates we can associate 
a unique admissible collection  of aggregates. 
 
For an admissible collection of aggregates 
$\{w'_1, \dots, w'_n \}_{\text{adm}}$, one has the constraint
$\sum_{i=1}^{n}h(w'_i)=0$. 
If we let, for  $i=1,\dots,n$,  $w_i$ denote  
the standard aggregates corresponding to $w'_i$, 
this constraint reads also 
$\sum_{i=1}^{n}h(w_i)=0$. 
From (\ref{fprime}), we then get:
\begin{equation}
\label{wallpf2}
  e^{- \beta F' ( \overline{V} )} 
= ( c^*_1 e^{- \beta J_1} + c^*_2 e^{- \beta
  J_2} )^N \sum_{\{ w_1, \ldots, w_n \}_{\text{adm}}} \prod_{i = 1}^n z ( w_i
  )\delta \Big( \sum_{i=1}
^{n}h(w_i)  ,0\Big)
\end{equation}
where the activities  $z(w)$ of aggregates are also given by (\ref{act}).

We now introduce the notion of \textit{elementary walls} which are the walls
corresponding to the elementary interfaces mentioned in the beginning of
this section.

In two dimensions, an elementary wall $W_{\text{el}}$ is a wall that contains
two pairs $\left\{ s_x, s_y \right\}$ and $\left\{ s_x, s_z \right\}$ such
that the site $x$ is occupied, the sites $y$ and $z$ are empty, $y$ is above
$x$ and $z$ is to the left or to the right of $x$.

The activity of  such a wall are easily computed and we
have
\begin{equation}
  \sum_{W_{\text{el}} : \text{supp} W_{\text{el}} = \left\{ x, y, z \right\}}
  z ( W_{\text{el}} ) =
\frac{c^*_1 e^{- 2 \beta J_1} + c^*_2 e^{- 2 \beta
     J_2}}{c^*_1 e^{- \beta J_1} + c^*_2 e^{- \beta J_2}}
\label{eq:ex1}
\end{equation}
In three dimensions, we call elementary wall, either a wall
(corresponding to the elementary interface $I_{\text{el}}^{\text{up}}$) that
contains four pairs $\left\{ s_x, s_{x_i} \right\}$, $i = 1, \ldots ., 4$,
such that the site $x$ is occupied, the sites $x_i$ are empty and all the
sites leave on the plane $x^3 = 0$, or a wall (corresponding to the elementary
interface $I_{\text{down}}^{\text{up}}$) that contains eight pairs $\left\{
s_x, s_{x_i} \right\}$, $\left\{ s_{x_i}, s_{y_i} \right\}, i = 1, \ldots .,
4$, such that the sites $x$ and $z_i$ are empty, the sites $y_i$ are occupied,
the sites $x$ and $x_i$ leaves on the plane $x^3 = - 1$, the sites $y_i$
leaves on the plane $x^3 = 0$.

In the first case we have
\begin{equation}\label{ex2}
  \sum_{W_{\text{el}} : \text{supp} W_{\text{el}} = \left\{ x, x_1, x_2, x_3,
  x_4 \right\}} z ( w_{\text{el}} ) =\frac{c^*_1 e^{- 5 \beta J_1} + c^*_2 e^{-
     5 \beta J_2}}{c^*_1 e^{- \beta J_1} + c^*_2 e^{- \beta J_2}}
\end{equation}
while in the second case
\begin{equation}\label{ex3}
  \sum_{W_{\text{el}} : \text{supp} W_{\text{el}} = \left\{ x, x_1, x_2, x_3,
  x_4, y_1, y_2, y_3,
  y_4 \right\}} z ( w_{\text{el}} ) =
\frac{( c^*_1
     e^{- 2 \beta J_1} + c^*_2 e^{- 2 \beta J_2} )^4}{( c^*_1 e^{- \beta J_1} +
     c^*_2 e^{- \beta J_2} )^4}
\end{equation}

For non elementary walls and aggregates, the activities (\ref{act})  may be bounded
as
\begin{eqnarray}
  z ( w ) 
& \leq & 
\frac{( a_g e^{- \beta J} )^{L ( w )}}{( c^*_1 e^{- \beta
  J_1} + c^*_2 e^{- \beta J_2} )^{| \pi ( w ) |}}
\\
  \sum_{w : L ( w ) = L} z ( w ) 
& \leq & 
\frac{( a_g e^{- \beta J} )^{L}
}{( c^*_1
  e^{- \beta J_1} + c^*_2 e^{- \beta J_2} )^{| \pi ( w ) |}}
\end{eqnarray}
where
\[ a_g = \left\{ \begin{array}{clc}
     8 e ( e - 1 ) \kappa e^{- \alpha} e^{\beta J}& \text{if}\  L ( w ) \geq 2d & \\
     1 & \text{otherwise} & 
   \end{array} \right. 
\]
This allows to exponentiate the partitions
functions of the gas of aggregates in the RHS of (\ref{wallpf}) and (\ref{wallpf2}),
for low enough temperatures. 
We define
multi-indexes $Y$ corresponding to aggregates as function from the set of aggregates
into the set of non negative integers, and we let $\text{supp} \hspace{0.25em}
Y = \cup_{w : Y ( w ) \geq 1} \text{supp} \hspace{0.25em} w$. 
The truncated
functional corresponding to the activities $z$ is given by
\begin{equation}
  \Psi ( w ) = \frac{a ( Y )}{\prod_w Y ( w ) !} \prod_w z ( w )^{Y ( w )}
\end{equation}
where the factor $a ( Y )$ is defined as in (\ref{eq:fcomb}).
Then,
\begin{equation}
    \sum_{  
\{ w_1, \ldots, w_n \}_{\text{adm}}
} \prod_{i = 1}^n z (
  w_i ) 
= \exp 
\Big\{ 
\sum_{Y : \text{supp} \hspace{0.25em} Y \subset \overline{V}}
  \Psi ( Y )
\Big\}
  =
 e^{
 - \beta N \mathcal{F} + \sigma (  \overline{V} \mid \Psi )
}
\end{equation}
in $3$--dimensions and  
\begin{eqnarray}
 \sum_{  
\{ w_1, \ldots, w_n \}_{\text{adm}}
} \prod_{i = 1}^n z (
  w_i )  
\delta \Big( \sum_{i=1}^{n}h(w_i)  ,0\Big)
&=&
\Pr
\Big\{
\sum_{i=1}^{n}h(w_i)  
 =0
\Big\}  
\exp 
\Big\{ 
\sum_{Y : \text{supp} \hspace{0.25em} Y \subset \overline{V}}
  \Psi ( Y )
\Big\}
\nonumber
\\
 & =&
\Pr \Big\{ \sum_{i=1}^{n}h(w_i)    =0 \Big\} 
e^{
 - \beta N \mathcal{F} + \sigma (  \overline{V} \mid \Psi )
}
\end{eqnarray}
in $2$--dimensions.
Here
\begin{equation}
  - \beta \mathcal{F} =  \sum_{Y : \text{supp} Y \ni 0} 
\frac{
\Psi (Y )
}{
| \text{supp} \hspace{0.25em} Y \cap \Pi_0   |} 
  \label{eq:series2}
\end{equation} 
\begin{equation}\label{bord}
  \sigma (  \overline{V} \mid \Psi ) 
 = 
 -
\sum_{Y:\text{supp} \hspace{0.25em} Y \cap\overline{V}^c\not= \emptyset } \Psi ( Y )
\frac{
| \text{supp} \hspace{0.25em} Y \cap \pi ( \overline{V} ) |
}{
| \text{supp} \hspace{0.25em} Y \cap \Pi_0   |}  
\end{equation}
and 
\begin{equation}
  \label{eq:proba}
  \Pr \Big\{   \sum_{i=1}^{n}h(w_i)  =0 \Big\} =
\frac{
\sum_{  
\{ w_1, \ldots, w_n \}_{\text{adm}}
}
 \prod_{i = 1}^n z (  w_i )
 \delta \Big( \sum_{i=1}^{n}h(w_i)  ,0\Big)
}{ 
\sum_{  
\{ w_1, \ldots, w_n \}_{\text{adm}}
}
 \prod_{i = 1}^n z (
  w_i ) }
\end{equation}
The series (\ref{eq:series2}) converges whenever
$ 8 e(e-1) \kappa^2 e^{- \alpha}<1$.
This can be seen  by verifying the convergence condition (\ref{eq:condconf})
with the activities $z(w)$ and with the contours replaced by aggregates.

Notice that the multi-indexes involved in the sum of (\ref{bord})
 intersects both $\pi (  \overline{V})$ and 
$ \overline{V}^c =\mathbb{Z}^d \setminus
 \overline{V}$. 
Thus at low temperatures, this sum can be bounded by $L^{d - 2}$ times a
constant so that this term will 
give no contribution to the surface tension in the thermodynamic limit.
In addition,
the probability $ \Pr\{   \sum_{i=1}^{n}h(w_i)  =0\} $,
may be controlled by known techniques  (see e.g. \cite{G,BF,DKS,DKLR}) and
$$\lim_{L \to \infty} (1/N) \ln \Pr\{   \sum_{i=1}^{n}h(w_i)  =0\}=0 $$
Hence, by taking the thermodynamic limit $L\to \infty$ in equations 
(\ref{wallpf}) and (\ref{wallpf2}) we obtain:
\begin{equation}
 e^{- \beta ( \tau_{\text{mixt}, \text{emp}} -
   \mathcal{F} )} = c^*_1 e^{- \beta J_1} + c^*_2 e^{- \beta J_2} 
\end{equation}

Whenever a multi-index $Y$ contains only one aggregate $w$ ($Y (
w ) = 1$ and $Y ( w' ) = 0$ for $w' \neq w$) one has
$\Psi ( Y ) = z ( w )$.
The leading terms of $\mathcal{F}$ are then obtained with 
the help of relations
(\ref{eq:ex1})--(\ref{ex3}). This gives the expressions (\ref{eq:free1}) and (\ref{eq:free2}). 
Taking furthermore  into account the expression (\ref{eq:tjf})  
ends the proof of the theorem.

\subsection*{Acknowledgments}

J.\ R.\ thanks the Centre de Recherche en Mod\'elisation 
Mol\'eculaire, Universit\'e de Mons--Hainaut 
for warm hospitality and financial support.

\bibliographystyle{unsrt}

\begin{thebibliography}{10}

\bibitem{Gu}
{E.\ A.\ Guggenheim}.
\newblock {\em Trans.\ Faraday Soc.}, {\bf 41}:150, 1945.

\bibitem{DP}
{R.\ Defay and I.\ Prigogine}.
\newblock {\em Trans.\ Faraday Soc.}, {\bf 46}:199, 1950.

\bibitem{Eb}
{J.\ G.\ Eberhart}.
\newblock {\em J.\ Phys.\ Chem.}, {\bf 70}:1183, 1966.

\bibitem{Sz}
{B.\ von Szyszkowsky}.
\newblock {\em J.\ Phys.\ Chem.}, {\bf 64}:385, 1908.

\bibitem{Sz2}
{H.\ P.\ Meissner and A.\ S.\ Michaels}.
\newblock {\em Ind.\ Eng.\ Chem.}, {\bf 42}:2782, 1949.

\bibitem{Ad}
{A.\ W.\ Adamson}.
\newblock {\em Physical Chemistry of Surfaces}.
\newblock John Wiley and Sons, New--York, 1997.

\bibitem{DPBE}
{R.\ Defay, I.\ Prigogine, A.\ Bellmans, and D.\ H Everett}.
\newblock {\em Surface tensions and absorption}.
\newblock Longmans, Green and Co., London, 1966.

\bibitem{DR}
{J.\ De Coninck and J.\ Ruiz}.
\newblock {Interfacial tensions for binary mixtures versus Solid--On--Solid
  models}.
\newblock {\em Preprint}, 2003.

\bibitem{WW}
{J.\ C.\ Wheeler and B.\ Widom}.
\newblock {\em J.\ Chem.\ Phys.}, {\bf 52}:5334, 1970.

\bibitem{LG}
{J.\ L. Lebowitz and G.\ Gallavotti}.
\newblock {Phases transitions in binary lattice gases}.
\newblock {\em J.\ Math.\ Phys.}, {\bf 12}:1129, 1971.

\bibitem{G}
{G.\ Gallavotti}.
\newblock Phase separation line in the two--dimensional {Ising} model.
\newblock {\em Commun.\ Math.\ Phys.}, {\bf 27}:103, 1972.

\bibitem{D1}
{R.\ L.\ Dobrushin}.
\newblock Gibbs states describing the coexistence of phases for a
  three-dimensional {Ising} model.
\newblock {\em Theory Prob.\ Appl.}, {\bf 17}:582, 1972.

\bibitem{BEG}
{M.\ Blume, V.\  Emery, and R.\ B.\  Griffiths}.
\newblock {\em Phys.\ Rev.\ A}, {\bf 4}:1071, 1971.

\bibitem{R1}
D.\ Ruelle.
\newblock {\em Statistical Mechanics: Rigorous Results}.
\newblock Benjamin, New--York, Amsterdam, 1969.

\bibitem{R2}
D.\ Ruelle.
\newblock {\em Thermodynamic Formalism}.
\newblock Addison Wesley, Reading, 1978.

\bibitem{S}
{Ya.\ G.\ Sinai}.
\newblock {\em Theory of Phase Transitions: Rigorous Results}.
\newblock Pergamon Press, London, 1982.

\bibitem{Z}
{M.\ Zahradnik}.
\newblock {An alternate version of {Pirogov}--{Sinai} theory}.
\newblock {\em Commun.\ Math.\ Phys.}, {\bf 93}:359, 1984.

\bibitem{GMM}
{A.\ Martin L\"{o}f, G.\ Gallavotti, and S.\ Miracle--Sole}.
\newblock {Some problem connected with the coexistence of phase in the {Ising}
  model}.
\newblock In {\em Statistical mechanics and mathematical problems, Lecture
  Notes in Physics}, volume~20, page 162. Springer, Berlin, 1973.

\bibitem{M}
S.\ Miracle-Sol\'{e}.
\newblock On the convergence of cluster expansion.
\newblock {\em Physica}, {\bf A 279}:244, 2000.

\bibitem{KP}
{R.\ Kotecky, and D.\ Preiss}.
\newblock Cluster expansion for abstract polymer models.
\newblock {\em Commun.\ Math.\ Phys.}, {\bf 103}:491, 1986.

\bibitem{BLP2}
{J.\ Bricmont, J.\ L.\ Lebowitz, and C.--E.\ Pfister}.
\newblock On the surface tensions of lattice systems.
\newblock {\em Ann.\ N.\ Y.\ Acad.\ Sci.}, {\bf 337}:214, 1980.

\bibitem{BLP}
{J.\ Bricmont, J.\ L.\ Lebowitz, and C.--E.\ Pfister}.
\newblock Non-translation invariant {Gibbs} states with coexisting phases
  {III}.
\newblock {\em Commun.\ Math.\ Phys.}, {\bf 66}:267, 1979.

\bibitem{Mi}
S.\ Miracle-Sol\'{e}.
\newblock Surface tension, step free energy, and facets in the equilibrium
  crystal.
\newblock {\em J.\ Stat.\ Phys.}, {\bf 79}:183, 1995.

\bibitem{A}
{D.\ B.\ Abraham}.
\newblock Surface structures and phase transitions--{Exact} results.
\newblock In {\em Phase Transitions and Critical Phenomena, {\rm volume 10}}.
  C.\ Domb and J.\ L.\ Lebowitz, editors, Academic Press, London, 1986.

\bibitem{P}
{C.--E.\ Pfister}.
\newblock Interface and surface tension in {Ising} model.
\newblock In {\em Scaling and Self--similarity in Physics, {\rm J.\ Fr\"ohlich,
  editor}}. Birkauser, 1982.

\bibitem{Temp}
{H.\ N.\ V.\ Temperley}.
\newblock {\em Proc.\ roy.\ Soc. \bf{A}}, {\bf 199}:3, 1949.

\bibitem{Progress}
 Work in progress.


\bibitem{BI}
{C.\ Borgs and J.\ Imbrie}.
\newblock A unified approach to phase diagrams in fields theory and statistical
  mechanics.
\newblock {\em Commun.\ Math.\ Phys.}, {\bf 123}:305, 1989.

\bibitem{DLR}
{C.\ Dobrovolny, L.\ Laanait, and J.\ Ruiz}.
\newblock Surface transitions of the semi--infinite {Potts} model {I}: the high
  bulk temperature regime.
\newblock {\em J.\ Stat.\ Phys.}, {\bf 114}:1269, 2004.

\bibitem{BF}
{J.\ Bricmont and J.\ Fr\"ohlich}.
\newblock Statistical mechanical methods in particle structure analysis of
  lattice field theories {II}.
\newblock {\em Commun.\ Math.\ Phys.}, {\bf 98}:553, 1985.

\bibitem{DKS}
{R.\ L.\ Dobrushin, R.\ Kotecky, and S.\ B.\ Shlosman}.
\newblock {\em {Wulff Construction: A Global Shape from Local Interactions}}.
\newblock Providence, New--York, 1992.

\bibitem{DKLR}
{J.\ De Coninck, R.\ Kotecky, L.\ Laanait, and J.\ Ruiz}.
\newblock {SOS} approximants for {Potts} crystal shapes.
\newblock {\em Physica}, {\bf A 189}:616, 1992.

\end{thebibliography}

\end{document}